\documentclass[twocolumn,aps,showpacs,prb,tightenlines,amsmath,amssymb]{revtex4}
\usepackage{bm}
\usepackage{graphicx}              
\usepackage{amssymb}               
\usepackage{amsmath}                     
\usepackage{colordvi}
\usepackage{calrsfs} 
\usepackage{mathrsfs}
\usepackage[thinlines,thiklines]{easybmat}
\newcommand{\bgreek}[1]{\mbox{\boldmath$#1$\unboldmath}}
\begin{document}   

\title{Gapped triplet $p$-wave superconductivity in strong
  spin-orbit-coupled semiconductor quantum wells in
proximity to $s$-wave superconductor}
 
\author{T. Yu}
\author{M. W. Wu}
\thanks{Author to whom correspondence should be addressed}
\email{mwwu@ustc.edu.cn.}
\affiliation{Hefei National Laboratory for Physical Sciences at Microscale, Department of Physics,
and CAS Key Laboratory of Strongly-Coupled Quantum Matter Physics,
University of Science and Technology of China, Hefei, Anhui, 230026,
China} 
\date{\today}

\begin{abstract} 
We show that the {\it gapped} triplet superconductivity, i.e., a
triplet superconductor with triplet order parameter, can be realized in 
strong spin-orbit-coupled (100) quantum wells in
proximity to $s$-wave superconductor. It is revealed that with the singlet order
parameter induced from the superconducting proximity effect,
 in quantum wells, not only can the triplet pairings arise
due to the spin-orbit coupling, but also the triplet order parameter can be
induced due to the repulsive effective electron-electron interaction,
including the electron-electron Coulomb and electron-phonon interactions. This
is a natural extension of the work of de Gennes, in which the
repulsive-interaction-induced
 singlet order parameter arises in the normal metal in
proximity to $s$-wave superconductor [Rev. Mod. Phys. {\bf 36}, 225 (1964)].
 Specifically, we derive the effective Bogoliubov-de
Gennes equation, in which the self-energies due to the effective electron-electron
interactions contribute to the singlet and triplet order parameters. It is
further shown that for the singlet order parameter, it is efficiently suppressed due to this
self-energy renormalization;
 whereas for the triplet order parameter, it is the $p$-wave
 ($p_x\pm ip_y$) one with the ${\bf
   d}$-vector parallel to the effective magnetic field
 due to the spin-orbit coupling. Finally, we perform
the numerical calculation in InSb (100) quantum
wells. Specifically, we reveal that the Coulomb interaction is much more important
than the electron-phonon interaction at low temperature. Moreover, it shows that
with proper electron density, the minimum of the renormalized singlet and the
maximum of the induced triplet order parameters are
comparable, and hence can be experimentally distinguished. 
\end{abstract}

\maketitle
\section{Introduction}
In recent years, triplet
 superconductivity has attracted much
 attention, which provides the possibility to realize the nondissipative
  spin transport and hence has potential application
 in spintronics.\cite{Spin_tunneling,Proximity_S_F,Triplet_S_F,SF1,Supercurrents,Superconducting_spintronics,
Awschalom,Zutic,fabian565,wu-review,notebook}
To confirm or realize the triplet superconductivity, much 
efforts have been made to several potential systems, including the
unconventional superconductor ${\rm Sr_2RuO_4}$,\cite{SrRuO_impurity,SrRuO_spin_lattice,
SrRuO_triplet,SrRuO_triplet2,SrRuO_triplet3,Lin1,Lin7,Lin8}
 the conventional superconductor-ferromagnet (S-F)
  interface with induced odd-frequency and even-momentum triplet
pairings,\cite{Proximity_S_F,Triplet_S_F,SF1,Linder1,
Tokatly_PRLB,exp_realization1,exp_realization2,SOC_S_F} conventional
 superconductors with induced spin-orbit coupling (SOC) in
  the surface or interface which possess even-frequency and odd-momentum triplet
pairings,\cite{Gorkov_Rashba,Mes_Chan,long_helix,Maslov_transverse,semiconductor,SOC_S_F,Linder_SOC}
 and the non-centrosymmetric
superconductors including the heavy fermion system.\cite{Sigrist,Sigrist_PRL,Sigrist_book}

Specifically, ${\rm Sr_2RuO_4}$ was
suggested to be the triplet $p$-wave
superconductor which may arise from the spin-fluctuation-induced attractive
potential between the triplet
states,\cite{Fay,Anderson_PRL,Anderson_PRB,SrRuO_triplet}
 whose experimental confirmation
 is still in progress.\cite{SrRuO_impurity,SrRuO_spin_lattice,
SrRuO_triplet,SrRuO_triplet2,SrRuO_triplet3,Lin1,Lin7,Lin8}
 In conventional S-F interface, it is shown that with the existence of the exchange field due
  to the ferromagnet, the spin-degeneracy is lifted. Accordingly, the 
odd-frequency and even-momentum triplet Cooper pairings emerge at the interface of
  S-F, in which the triplet order parameter
  is further shown to be zero.\cite{Proximity_S_F,Triplet_S_F,SF1,Linder1,Tokatly_PRLB,
exp_realization1,exp_realization2,Berezinskii} One notes that the
  order parameter directly contributes to the superconducting gap. Specifically, it is well established that
 with the inhomogeneous ferromagnet, the induced triplet pairs can
diffuse into the ferromagnetic materials over distances
much larger than the singlet ones, which is referred to as long-range proximity
effect.\cite{Tokatly_PRLB}
Similar to the exchange field, SOC can also lift the spin degeneracy and
 hence provides another possibility to realize the
  triplet superconductivity.\cite{Gorkov_Rashba,Mes_Chan,long_helix,Maslov_transverse,semiconductor}
 This possibility was first pointed out by Gor'kov and
Rashba when considering the $s$-wave superconductivity with the SOC induced by
the absorption of ion.\cite{Gorkov_Rashba} It was shown that due to the lift of the spin-degeneracy
by the SOC, mixed singlet-triplet Cooper pairings arise in which the triplet
part is odd-momentum and even-frequency. We point out here that with the
momentum-independent attractive potential between electrons from the $s$-wave channel,
 no triplet gap or triplet order
parameter arises in the
superconductor. As a natural extension, much efforts are focused on the
  system with the SOC in proximity to the $s$-wave superconductors, including
  spin-orbit-coupled metals\cite{Tokatly_PRLB,Linder_SOC,Mes_Chan,long_helix,Maslov_transverse} and
  even semiconductors.\cite{semiconductor}
 Finally, in the non-centrosymmetric
superconductor, with the SOC naturally existing in the superconductor itself,
 it is shown that if proper forms of {\it attractive} effective
electron-electron (e-e) interaction potential are
realized from the symmetry analysis, triplet gap or triplet order parameter can
be realized.\cite{Sigrist,Sigrist_PRL,Sigrist_book} 
 Specifically, it is shown that when the ${\bf d}$-vector of the triplet order
parameter is parallel to the effective magnetic field due
to the SOC, the system can have minimum free energy.\cite{Sigrist_PRL}

In above systems, one can see that in ${\rm Sr_2RuO_4}$ and non-centrosymmetric
superconductors, the triplet order parameter can naturally arise from
proper effective
 e-e interaction
  potential.\cite{Sigrist,Sigrist_PRL,SrRuO_triplet}
 However, their experimental confirmations are still
 in progress.\cite{Sigrist,Sigrist_PRL,SrRuO_impurity,SrRuO_spin_lattice,
SrRuO_triplet,SrRuO_triplet2,SrRuO_triplet3,Lin1,Lin7,Lin8}
In contrast to this, the triplet superconductivity in the system with the Zeeman
field or SOC,\cite{Tokatly_PRLB,Linder_SOC,Mes_Chan,long_helix,
Maslov_transverse,semiconductor,Proximity_S_F,Triplet_S_F,SF1,Linder1,
exp_realization1,exp_realization2} which is in proximity to the $s$-wave superconductor, is
relatively easy to be realized and manipulated with the flexible manipulation of
the strength and type of the SOC.\cite{Rashba1,Rashba2,Dresshaus} Specifically,
the triplet superconductivity in conventional S-F interface or the S-F-S
Josephson junction has been experimentally confirmed by observing the structure
of the energy gap\cite{exp_realization1,exp_realization2}
or $0\mbox{-}\pi$ transition of Josephson
effect.\cite{0_pi_origin,0_pi_theory2,0_pi_experiment,0_pi_Linder}
 Nevertheless, it is shown that 
although there exists triplet pairings, no triplet order parameter arises in both the
interface of conventional S-F and the system with SOC in proximity to conventional
$s$-wave superconductors.\cite{Gorkov_Rashba,Proximity_S_F,Triplet_S_F,SF1}
 As a consequence, the elementary excitation spectra cannot be
influenced by the triplet pairings, and are only determined by the
 singlet order parameter. Furthermore, in the interface of conventional S-F, in
 the ferromagnet side, when the singlet order parameter can be  
 neglected due to the weak interaction potential, the system shows gapless
 structure.\cite{Proximity_S_F,Triplet_S_F,exp_realization1,exp_realization2,D_Gennes}  
 Due to the gapless structure, the experimental realization of the gapless triplet superconductivity is
 performed at extremely low
 temperature due to the absence of the gap protection. 

One further notes that in above {\it metal} systems with the Zeeman field or SOC
which are in proximity to the
$s$-wave superconductors, the e-e interaction can be neglected due to
the strong screening.\cite{Proximity_S_F,Triplet_S_F,SF1} Nevertheless, in the study of the boundary effects
 in superconductors-normal metal, de Gennes pointed out that the
Cooper pairs tunneling or diffusing from the $s$-wave superconductor also experience the many-body interaction in the
normal metal, in which the singlet order parameter can be induced even with
a {\em repulsive} effective e-e interaction.\cite{D_Gennes} Accordingly, 
it is natural to consider the possibility to realize the triplet order
 parameter in system with the Zeeman field or SOC in proximity to the
 $s$-wave superconductor from the effective e-e interaction,
 which can protect the ground state and is promising to
 provide rich physics
 especially for the elementary excitation. As expected, this effect is
 significant only when the effective e-e interaction is strong.  This can be
 realized in low-dimensional 
 semiconductors with weak screening effect, based on the facts that the proximity-induced
 superconductivity from $s$-wave superconductor in two-dimensional (2D) electron gas, including
 InAs\cite{2D_super_InAs1,2D_super_InAs2} and GaAs\cite{GaAs28,GaAs29,2D_super_GaAs}
   heterostructures, and quantum nanowire\cite{wire1,wire2,wire3,wire4} has been reported in the
   literature. 
 
In the present work, we show that a {\it gapped} triplet superconductivity with triplet order
   parameter can be realized in 2D electron gas of the spin-orbit-coupled 
   quantum wells (QWs) in proximity to the
   $s$-wave superconductor. This triplet order
   parameter is induced by the effective e-e interaction including the
   Coulomb and electron-phonon (e-p)
interactions, even the total interactions are repulsive.
 Specifically, based on the superconducting proximity effect,
 it has been shown that the singlet order parameter can be induced in
   the 2D electron gas.\cite{Dar_Sarma1,Dar_Sarma2,Dar_Sarma3} With this 
proximity-induced singlet order parameter, it can be further
shown that the triplet pairings are induced due to the SOC.\cite{Gorkov_Rashba} 
Furthermore, we derive the effective Bogoliubov-de
Gennes (BdG)
 equation, in which the self-energy due to the {\it momentum-dependent} e-e and e-p interactions is
 presented explicitly.\cite{Fetter,Mahan,Abrikosov} Specifically, in
 the effective BdG equation, the 
self-energy from the effective e-e interaction leads to the $p$-wave
($p_x\pm ip_y$) triplet
order parameter. Moreover, from the effective BdG equation, it is discovered
that the proximity-induced singlet order
parameter is also inevitably renormalized by the effective e-e interactions.

To make the physics clearer, we further carry out the numerical
  calculation in the specific material InSb (100) QWs,
 in which there exists strong SOC.\cite{material_book,parameter_SOC} The
  calculations show that the self-energy due to the e-p interaction is much
  smaller than that due to the e-e Coulomb interaction at low temperature ($T=$2~K), and hence only the
  Coulomb interaction needs to be considered here. For the
  renormalized singlet order parameter, it is always smaller than the
  proximity-induced one, as the renormalization from the Coulomb
  interaction is in the opposite sign against the proximity-induced
  order parameter. Moreover, it is
  shown that it only depends on the
  magnitude of the momentum, and decreases with the
  increase of the energy due to the suppression of the Coulomb interaction at
  high energy. For the induced triplet order parameter, it  
  depends not only on the magnitude of the momentum,
 but also on its angle. Specifically,
 in the momentum-module dependence, a {\it peak}
  is predicted from our theory. In the angular dependence, it is proved that
 the ${\bf d}$-vector of this triplet order parameter
is parallel to the effective magnetic field due to the SOC, and hence is protected
by the SOC.\cite{Sigrist_PRL} Finally, we study the electron density dependencies
of the singlet and triplet order parameters in detail. Rich behaviors arise when
the electrons populate different energy bands, which are split by the strong
SOC. It is further found that with proper electron
density ($n_e\approx 8\times 10^{14}$~cm$^{-2}$),
 the minimum of the renormalized singlet order parameter and the
 maximum of the induced triplet order parameter are
comparable, which provides an ideal condition to observe and distinguish these order parameters
in the experiment.

This paper is organized as follows. In Sec.~{\ref{Model}}, we set up the
  model and Hamiltonian. In Sec.~{\ref{analytical}}, we present the
  analytical results including the effective BdG equation (Sec.~{\ref{BdG}})
  and the calculation of the e-e and e-h self-energies
  (Sec.~\ref{self_energy}). In Sec.~{\ref{numerical}}, the numerical results are
  performed in InSb (100) QWs, in which both the suppression of the
  proximity-induced singlet order parameter (Sec.~\ref{SS}) and induced triplet order
  parameter (Sec.~\ref{PP})
  are discussed. We summarize and discuss in Sec.~{\ref{summary}}.

\section{Model and Hamiltonian}
\label{Model}
We start our investigation from the Hamiltonian of the (100) {\it symmetric} QWs in proximity to the $s$-wave
superconductor, which is composed by the Hamiltonian of (100) QWs
$\hat{H}_{\rm QW}$ (Sec.~\ref{H_QWs}) and the Hamiltonian of the $s$-wave
superconductor $\hat{H}_{\rm S}$ (Sec.~\ref{H_superconductor}).

\subsection{Hamiltonian of (100) QWs}
\label{H_QWs}
The Hamiltonian of (100) QWs is written as  
\begin{equation}
\hat{H}_{\rm QW}=\hat{H}_{\rm QW}^{\rm k}+\hat{H}_{\rm QW}^{\rm soc}+\hat{H}_{\rm QW}^{\rm
  ee}+\hat{H}_{\rm QW}^{\rm ep}.
\label{Hamiltonian}
\end{equation}
Here, $\hat{H}_{\rm QW}^{\rm k}$, $\hat{H}_{\rm QW}^{\rm soc}$, $\hat{H}_{\rm QW}^{\rm
  ee}$ and $\hat{H}_{\rm QW}^{\rm ep}$ are the kinetic energy of the electron,
the SOC, the e-e Coulomb interaction and e-p interaction, respectively. In QWs,
 by using the field operator defined in Nambu$\otimes$spin space $\hat{\Psi}({\bf
  r})=\big(\psi_{\uparrow}({\bf r}),\psi_{\downarrow}({\bf
  r}),\psi_{\uparrow}^{\dagger}({\bf r}),\psi_{\downarrow}^{\dagger}({\bf r})\big)^T$, these Hamiltonians are
given as follows. The kinetic
energy reads ($\hbar\equiv 1$ throughout this paper)
\begin{equation}
\hat{H}_{\rm QW}^{\rm k}=\frac{1}{2}\int d{\bf
  r}\hat{\Psi}^{\dagger}({\bf r})\big(\varepsilon_{\bf k}-\mu\big)\hat{\tau}_3\hat{\Psi}({\bf r}),
\end{equation}
where $\varepsilon_{\bf k}={\bf k}^2/(2m^*)$ with ${\bf k}=(k_x,k_y)$ being the momentum of electron, $m^*$ denotes
 the effective mass of electron, $\mu$ represents the chemical potential, 
and $\hat{\tau}_3={\rm diag}(1,1,-1,-1)$. 
The SOC Hamiltonian is
\begin{equation}
\hat{H}_{\rm QW}^{\rm soc}=\frac{1}{2}\int d{\bf
  r}\hat{\Psi}^{\dagger}({\bf r})\left(
\begin{array}{cc}
\hat{h}_{\rm soc}(\hat{\bf k})&0\\
0&\hat{h}_{\rm soc}^{*}(-\hat{\bf k})
\end{array}\right)\hat{\tau}_3\hat{\Psi}({\bf r}).
\label{SOC}
\end{equation}
Here, $\hat{h}_{\rm soc}({\bf k})=-\alpha
  \hat{k}_x\sigma_x+\alpha \hat{k}_y\sigma_y$ in which $\alpha=\gamma_D(\pi/a)^2$
  for the infinitely deep well 
  with $\gamma_D$ and $a$ being the Dresselhaus coefficient and well width,
  respectively, and ${\bgreek \sigma}=(\sigma_x,\sigma_y,\sigma_z)$
  are the Pauli matrices.\cite{e_p_formula2,JiangLei}

The e-e Hamiltonian is
  written as 
\begin{equation}
\hat{H}_{\rm QW}^{\rm ee}=\frac{1}{2}\int d{\bf r}d{\bf r}'V({\bf r}-{\bf
  r}')\Big[\hat{\Psi}^{\dagger}({\bf r})\hat{\tau}_3\hat{\Psi}({\bf r})\Big]
\Big[\hat{\Psi}^{\dagger}({\bf r}')\hat{\tau}_3\hat{\Psi}({\bf r}')\Big],
\end{equation}
where $V({\bf r}-{\bf
  r}')$ denotes the screened Coulomb potential, whose Fourier transformation is
represented by 
 $V({\bf k})=\frac{\displaystyle V_0({\bf k})}{\displaystyle
    1-P^{(1)}({\bf k})V_0({\bf k})}$.
 Here, $V_0({\bf k})=\displaystyle \int dy\frac{1}{\pi
      a}\frac{e^2}{\varepsilon_0\kappa_0(k^2+4y^2/a^2)}|I(y)|^2$ with
    $\varepsilon_0$
 and $\kappa_0$ standing for the vacuum permittivity and
relative dielectric constant; 
 $|I(y)|^2=\frac{\displaystyle \pi^4\sin^2(y)}{\displaystyle (\pi^2-y^2)^2y^2}$
 representing the form factor;
$P^{(1)}({\bf k})$ denoting the longitudinal polarization function, whose expression is derived
based on the linear-response theory with
density-density correlation function (refer to Appendix.~\ref{BB}).\cite{Fetter,Mahan,Abrikosov,dielectric}

Finally, the e-p interaction Hamiltonian is denoted as
\begin{equation}
\hat{H}_{\rm QW}^{\rm ep}=\frac{1}{2}\int d{\bf r}d{\bf r}'g({\bf r}-{\bf
  r}')\hat{\Psi}^{\dagger}({\bf r})\hat{\tau}_3\Psi({\bf r})\phi({\bf r}'),
\end{equation}
where $g({\bf r}-{\bf
  r}')$ is the coupling potential between electron and phonon and $\phi({\bf
  r})$ is the phonon field operator. Specifically, at low temperature, we focus
on three electron-AC-phonon interactions due to the deformation potential
in LA-branch and piezoelectric coupling including LA and TA branches.
 The Fourier transformations of the coupling
potential $g({\bf r}-{\bf
  r}')$ between the AC phonons and electrons are explicitly given in Refs.~\onlinecite{e_p_formula1,e_p_formula2}.

\subsection{Hamiltonian of $s$-wave superconductor}
\label{H_superconductor}
In the conventional $s$-wave superconductor, with the field operator in
 Nambu$\otimes$spin space expressed as $\hat{\Phi}({\bf r})=\big(\phi_{\uparrow}({\bf
   r}),\phi_{\downarrow}({\bf r}),
\phi_{\uparrow}^{\dagger}({\bf r}),\phi_{\downarrow}^{\dagger}({\bf r})\big)^T$,
 the Hamiltonian is
expressed as
\begin{equation}
\hat{H}_{\rm S}=\frac{1}{2}\int d{\bf r}\hat{\Phi}^{\dagger}({\bf r})
\hat{H}_{\rm S}^{\rm BdG}\hat{\Phi}({\bf r}),
\end{equation}
where the BdG Hamiltonian $\hat{H}_{\rm S}^{\rm BdG}$ is written as
\begin{equation}
\hat{H}_{\rm S}^{\rm BdG}=\left(
\begin{array}{cccc}
\frac{\hat{\displaystyle {\bf p}}^2}{\displaystyle 2\tilde{m}}-\tilde{\mu} & 0 &0&-\Delta_0\\
0&\frac{\hat{\displaystyle
    {\bf p}}^2}{\displaystyle 2\tilde{m}}-\tilde{\mu}&\Delta_0&0\\
0&-\Delta_0^*&\frac{\hat{\displaystyle {\bf p}}^2}{\displaystyle 2\tilde{m}}-\tilde{\mu}&0\\
\Delta_0^*&0&0&\frac{\hat{\displaystyle {\bf p}}^2}{\displaystyle 2\tilde{m}}-\tilde{\mu}
\end{array}
\right)\hat{\tau}_3.
\label{S_BdG}
\end{equation}
In Eq.~(\ref{S_BdG}), for the electron in
the superconductor, ${\bf p}=(p_x,p_y,p_z)$ is the momentum; $\tilde{m}$ and $\tilde{\mu}$ are the mass and chemical
potential, respectively; and $\Delta_0$ denotes
the singlet gap, which is taken to be real in this work.

\section{Analytical results}
\label{analytical}

\subsection{Effective BdG equation in QWs}
\label{BdG}
In this section, we derive the effective BdG equation in QWs 
by using the equilibrium Green
function method in the Matsubara
  representation in the Nambu$\otimes$spin space, from which we obtain that
 the self-energies due to the
effective e-e interactions act as the effective singlet
 and triplet pairing potentials (order parameters).\cite{Fetter,Mahan,Abrikosov}
 Here, the zeroth-order Hamiltonian is chosen
to be $\hat{H}_{\rm QW}^0=\hat{H}_{\rm QW}^{\rm k}+\hat{H}_{\rm
  QW}^{\rm soc}$, and $\hat{H}_{\rm QW}^{\rm
  ee}$ and $\hat{H}_{\rm QW}^{\rm ep}$ are treated as perturbations.
In the Nambu$\otimes$spin space, the equilibrium Green function is defined as
\begin{equation}
G_{12}=-\hat{\tau}_3\langle T_{\tau}\hat{\Psi}_1\hat{\Psi}^{\dagger}_2\rangle,
\label{definition}
\end{equation} 
with $T_{\tau}$ representing the chronological product, $(1)=(\tau_1,{\bf r}_1)$
representing the imaginary$\mbox{-}$time--space point, and
$\langle \cdots\rangle$ denoting the ensemble
average.\cite{Fetter,Mahan,Abrikosov}

When no interactions are included, the eigenstates of $\hat{H}_{\rm
    QW}^0$ are expressed as the spinor wavefunction $U_n({\bf
  r})=\big(u_{n,\uparrow}({\bf r}),u_{n,\downarrow}({\bf r}),v_{n,\uparrow}({\bf
  r}),v_{n,\downarrow}({\bf r})\big)^T$ for the $n$th-state with eigen-energy
$E_n$, i.e., $\hat{H}_{\rm QW}^0U_n({\bf r})=E_nU_n({\bf r})$.
 For these eigenstates, the orthonormal
conditions
$\int d{\bf r}U_{n}^{\dagger}({\bf r})U_{n'}({\bf r})=\delta_{nn'}$ and
  $\sum_nU_{n}({\bf r})U_{n}^{\dagger}({\bf r}')=\delta({\bf r}-{\bf r}')$ are satisfied.
 Accordingly, the field operator in the Heisenberg
representation can be expanded by these eigenstates as
 ${\Psi}^{\dagger}(\tau,{\bf r})=\sum_ne^{E_n \tau}U_{n}^{\dagger}({\bf r})\alpha_n^{\dagger}$, where
$\alpha_n^{\dagger}$ is the creation operator for $n$th state. Accordingly, from
Eq.~(\ref{definition}), the
free Green function is represented as
\begin{eqnarray}
\nonumber
&&G_{12}^0=-\sum_n\hat{\tau}_3U_{n}({\bf r}_1)U_{n}^{\dagger}({\bf
  r}_2)e^{-E_n(\tau_1-\tau_2)}\\
&&\mbox{}\times\Big[\theta(\tau_1-\tau_2)\langle\alpha_n\alpha_n^{\dagger}\rangle
-\theta(\tau_2-\tau_1)\langle\alpha_n^{\dagger}\alpha_n\rangle\Big],
\label{zero_order}
\end{eqnarray}
where $\theta(\tau_1-\tau_2)$ is the Heaviside step function.
 Then from Eq.~(\ref{zero_order}),
one can obtain the dynamics equation for the free
Green function, 
\begin{equation} 
 (-\frac{\partial}{\partial{\tau_1}}\hat{\tau}_3-\hat{H}^{0}_{\rm
    QW}\hat{\tau}_3)G_{12}^{0}=\delta(1-2).
\label{Gorkov}
\end{equation}
In the Matsubara-frequency space, $G^0({\bf r}_1,{\bf r}_2;i\omega_m)
=\int_{0}^{\beta}d\tau e^{i\omega_m\tau}G^0({\bf r}_1,{\bf r}_2;\tau)$, where
$\beta=1/(k_BT)$ and
$\omega_m=(2m+1)\pi/\beta$ are Matsubara frequencies with $m$ being
  integer.
Then in this space, Eq.~(\ref{Gorkov}) is transformed into 
\begin{equation} 
 (i\omega_m\hat{\tau}_3-\hat{H}^{0}_{\rm
    QW}\hat{\tau}_3)G^0({\bf r}_1,{\bf r}_2; i\omega_m)=\delta({\bf r}_1-{\bf r}_2).
\label{BdG_frequency}
\end{equation}

When the interactions are considered, the eigenfunction
 and creation (annihilation) operator are expressed as $\tilde{U}_{n}$ and
$\tilde{\alpha}_n^{\dagger}$ ($\tilde{\alpha}_n$), respectively, whose
eigen-energy is $\tilde{E}_n$. Accordingly,
the Green function in Matsubara-frequency space is expressed as
\begin{equation}
G({\bf r}_1,{\bf r}_2;i\omega_m)=\sum_n\hat{\tau}_3\tilde{U}_{n}({\bf r}_1)\tilde{U}_{n}^{\dagger}({\bf
  r}_2)\frac{1}{i\omega_m-\tilde{E}_n}.
\end{equation} 
From the Dyson equation, one can also express the above Green
function as
\begin{eqnarray}
\nonumber
\hspace{-0.5cm}&&G({\bf r}_1,{\bf r}_2; i\omega_m)=G^0({\bf r}_1,{\bf r}_2; i\omega_m)+\int d{\bf
  r}_3d{\bf r}_4G^0({\bf r}_1,{\bf r}_3; i\omega_m)\\
\hspace{-0.5cm}&&\mbox{}\times\Sigma({\bf r}_3,{\bf r}_4; i\omega_m)G({\bf r}_4,{\bf r}_2; i\omega_m),
\label{Dyson}
\end{eqnarray}
where $\Sigma({\bf r}_3,{\bf r}_4; i\omega_m)$ are the self-energies due to
 $\hat{H}_{\rm QW}^{\rm
  ee}$ and $\hat{H}_{\rm QW}^{\rm ep}$. 
By performing the operation $(i\omega_m\hat{\tau}_3-\hat{H}^0_{\rm
    QW}\hat{\tau}_3)$ on Eq.~(\ref{Dyson}), with Eq.~(\ref{BdG_frequency}), one obtains
\begin{equation}
\hat{H}^0_{\rm QW}\tilde{U}_n({\bf r})+\int d{\bf r}'\Sigma({\bf r}-{\bf
  r}',i\omega_m)\hat{\tau}_3\tilde{U}_n({\bf r}')=\tilde{E}_n\tilde{U}_n({\bf r}).
\label{re_BdG}
\end{equation}
Specifically, in homogeneous space, Eq.~(\ref{re_BdG}) is written in the
momentum-space as
\begin{equation}
\big[\hat{H}^0_{\rm QW}({\bf k})+\Sigma({\bf
  k},i\omega_m)\hat{\tau}_3\big]\tilde{U}_n({\bf k})=\tilde{E}_n({\bf
  k})\tilde{U}_n({\bf k}).
\label{effective_imaginary}
\end{equation}
Finally, in the real-frequency space, by using the analytical continuation
  $i\omega_{m}\rightarrow \omega+i0^{+}$, Eq.~(\ref{effective_imaginary}) becomes
\begin{equation}
\big[\hat{H}^0_{\rm QW}({\bf k})+\Sigma({\bf
  k},\omega)\hat{\tau}_3\big]\tilde{U}_n({\bf k})=\tilde{E}_n({\bf
  k})\tilde{U}_n({\bf k}).
\label{effective}
\end{equation}

Eq.~(\ref{effective}) is the {\it effective BdG equation} in QWs, which can be
used to calculate the energy-spectra and wavefunction of the elementary excitation.
 Moreover, from the
structure of the self-energy $\Sigma({\bf
  k},i\omega_m)$, one can obtain the effective singlet and triplet order
parameters, which are presented in Sec.~\ref{singlet} and \ref{triplet}, respectively.

\subsection{Singlet and triplet order parameters from self-energy}
\label{self_energy}
In this part, we present the self-energies due to the superconducting
  proximity effect, e-e and e-p
interactions, respectively. One notes that the self-energies and
Green function should be calculated consistently, because the Green function determines the
self-energy and vice versa,  from Eq.~(\ref{Dyson}), the self-energy also 
influences the Green function. Therefore, when there exits two kinds of self-energies, i.e., the
self-energy due to the superconducting proximity effect and the self-energy due to e-e and e-p
interactions, their calculations are complicated because 
they are influenced on each other through the determination of the
Green function. However, when the two kinds of interactions are not comparable,
the calculation is highly simplified. Here, the e-e and e-p interactions are weaker than the
one due to superconducting proximity effect. This makes it reasonable to calculate the
self-energy due to the superconducting proximity effect 
without consideration of the e-e and e-p interactions, from which the Green
function including the
  superconducting proximity effect is determined (Sec.~\ref{singlet}). With this Green
function, we further calculate the self-energy due to e-e and e-p interactions (Sec.~\ref{triplet}).
  
\subsubsection{Self-energy and Green function due to superconducting proximity effect}
\label{singlet}
In this part, the self-energy and Green function due to the superconducting proximity effect
  are presented. Specifically, the self-energy due to the superconducting
  proximity effect
  is written as
\begin{equation}
\Sigma_s(\bar{k})\approx\left(
\begin{array}{cccc}
0 & 0 &0&-\Delta(\bar{k})\\
0&0&\Delta(\bar{k})&0\\
0&-\Delta^*(\bar{k})&0&0\\
\Delta^*(\bar{k})&0&0&0
\end{array}
\right),
\label{Self_tunneling_effective}
\end{equation} 
where $\bar{k}\equiv(i\omega_m,{\bf k})$. Here, from Eqs.~(\ref{effective_imaginary})
and (\ref{effective}), one observes that 
$\Delta(\bar{k})$ acts as the singlet
order parameter in the QWs, which is referred to as proximity-induced singlet
order parameter in this work. 
One notices that this self-energy [Eq.~(\ref{Self_tunneling_effective})]
 can be induced from the single-particle tunneling
  between QWs and superconductors (refer to
  Appendix~\ref{AA})\cite{Dar_Sarma1,Dar_Sarma2,Hubbard1} and
    other possibilities.\cite{Triplet_S_F,note}

From Eq.~(\ref{Self_tunneling_effective}), we calculate the Green function for
the 2D electron gas 
  in QWs with the proximity-induced singlet order parameter included, based
on the Dyson's equation in frequency-momentum space,
\begin{equation}
G(\bar{k})=G_0(\bar{k})+G_0(\bar{k})\Sigma_s(\bar{k})G(\bar{k}).
\end{equation} By expressing
\begin{equation}
G(\bar{k})=\left(
\begin{array}{cccc}
G_{\uparrow\uparrow}(\bar{k}) & G_{\uparrow\downarrow}(\bar{k})&F_{\uparrow\uparrow}(\bar{k})
&F_{\uparrow\downarrow}(\bar{k})\\
G_{\downarrow\uparrow}(\bar{k})&G_{\downarrow\downarrow}(\bar{k})
&F_{\downarrow\uparrow}(\bar{k})&F_{\downarrow\downarrow}(\bar{k})\\
F^{*}_{\uparrow\uparrow}(-\bar{k})&F^{*}_{\uparrow\downarrow}(-\bar{k})
&G^*_{\uparrow\uparrow}(-\bar{k})&G^*_{\uparrow\downarrow}(-\bar{k})\\
F^{*}_{\downarrow\uparrow}(-\bar{k})&F^{*}_{\downarrow\downarrow}(-\bar{k})
&G^*_{\downarrow\uparrow}(\bar{k})&G^*_{\downarrow\downarrow}(-\bar{k})
\end{array}
\right),
\label{Green_tot}
\end{equation}
one obtains the normal Green function
\begin{eqnarray}
\nonumber
\hspace{-0.4cm}&&\left(\begin{array}{cc}
G_{\uparrow\uparrow}(\bar{k}) & G_{\uparrow\downarrow}(\bar{k})\\
G_{\downarrow\uparrow}(\bar{k})&G_{\downarrow\downarrow}(\bar{k})\\
\end{array}
\right)
\label{GG}\\
\nonumber
\hspace{-0.4cm}&&\mbox{}=\frac{1}{2}\left(
\begin{array}{cc}
A_{+}(\bar{k})
+A_{-}(\bar{k}) 
& h_{\bf k}\big[A_{+}(\bar{k})-A_{-}(\bar{k})\big]\\ 
h^*_{\bf k}\big[A_{+}(\bar{k})-A_{-}(\bar{k})\big]&A_{+}(\bar{k})+A_{-}(\bar{k}) \\
\end{array}
\right),\\
\end{eqnarray}
and anomalous (Gor'kov's ) Green function\cite{Abrikosov}
\begin{eqnarray}
\nonumber
\hspace{-0.4cm}&&\left(
\begin{array}{cc}
F_{\uparrow\uparrow}(\bar{k}) & F_{\uparrow\downarrow}(\bar{k})\\
F_{\downarrow\uparrow}(\bar{k})&F_{\downarrow\downarrow}(\bar{k})\\
\end{array}
\right)
\label{FF}\\
\nonumber
\hspace{-0.4cm}&&\mbox{}=\frac{1}{2}\left(
\begin{array}{cc}
h_{\bf k}\big[B_{-}(\bar{k})-B_{+}(\bar{k})\big]
& B_{+}(\bar{k})+B_{-}(\bar{k})\\
-B_{+}(\bar{k})-B_{-}(\bar{k})&
 h^*_{\bf k}\big[B_{+}(\bar{k})-B_{-}(\bar{k})\big]\\
\end{array}
\right).\\
\end{eqnarray}
In Eqs.~(\ref{GG}) and (\ref{FF}),
 $A_{\pm}(\bar{k})=\frac{\displaystyle i\omega_m+\epsilon_{{\bf k}\pm}}{\displaystyle
  (i\omega_m)^2-\epsilon_{{\bf k}\pm}^2-|\Delta(\bar{k})|^2}$ and
 $B_{\pm}(\bar{k})=\frac{\displaystyle \Delta(\bar{k})}{\displaystyle
  (i\omega_m)^2-\epsilon_{{\bf k}\pm}^2-|\Delta(\bar{k})|^2}$ with
$\epsilon_{{\bf k}\pm}=\frac{\displaystyle k^2}{\displaystyle 2m^*}\pm \alpha
k-\mu=E_{k,\pm}-\mu$; $h_{\bf k}=-e^{i\phi_{\bf k}}$ with $\phi_{\bf k}$ being
the angle of the momentum.

Some features can be revealed from the normal and anomalous Green functions [Eqs.~(\ref{GG})
  and (\ref{FF})] when there exists the SOC. 
Specifically, from the off-diagonal terms of the normal Green function
 [Eq.~(\ref{GG})], one concludes that there always exists
correlation for the electron with different spins due to the SOC.
 From the anomalous Green function [Eq.~(\ref{FF})], there
exist anomalous correlations for the electrons not only with the same spins,
i.e., the off-diagonal terms in Eq.~(\ref{FF}),
 but also with different spins, i.e., the diagonal terms in
 Eq.~(\ref{FF}). Therefore, one can realize the mixed singlet-triplet pairings
 in spin-orbit-coupled QWs in proximity to conventional $s$-wave
 superconductors.\cite{Gorkov_Rashba} Nevertheless, from the effective BdG equation
 with the self-energy due to the superconducting proximity effect, i.e., Eq.~(\ref{effective}), one
 can see that only the singlet component
 contributes the order parameter. In the following, we can show that when the
 e-e and e-p interactions are further considered, the triplet order parameter is also
 induced and hence the gapped triplet superconductivity can be realized.

\subsubsection{Self-energy due to e-e and e-p interactions}
\label{triplet}
\begin{figure}[ht]
  {\includegraphics[width=8.15cm]{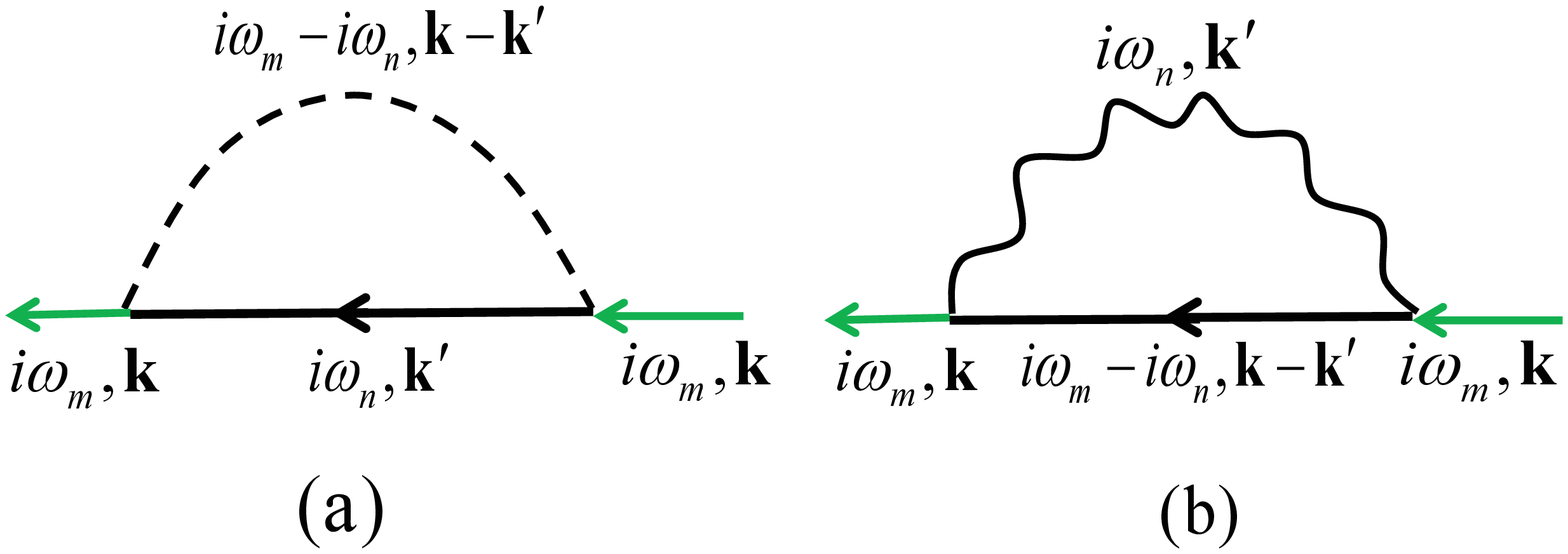}}
  \caption{(Color online) Feynman diagrams for the calculation of self-energies due to e-e [(a)]
    and e-p [(b)] interactions. Here, $\leftarrow$ represents the Green function
    $G(i\omega_m,{\bf k})$
 [Eq.~(\ref{Green_tot})] in matrix form. 
The black dashed curve in (a) and black wavy curve in (b) 
    represent the Coulomb potential and phonon Green function, respectively.}
  \label{figyw1}
\end{figure}

In this part, the self-energies due to e-e and e-p
  interactions are derived, whose Feynman diagram is shown in
  Figs.~\ref{figyw1}(a) and (b), respectively.
 For the e-e interaction, from the Feynman diagram in 
Fig.~\ref{figyw1}(a), the self-energy in the Matsubara
  representation is written as
\begin{equation}
\Sigma_{\rm ee}({\bf k})
=-\frac{1}{\beta}\int \frac{d{\bf k}'}{(2\pi)^2}V({\bf k}-{\bf
  k}')\sum_{n}G(i\omega_{n},{\bf k}').
\label{ee_finite}
\end{equation}
For the e-p
  interaction, from the Feynman diagram in Fig.~\ref{figyw1}(b), the self-energy reads
\begin{eqnarray}
\nonumber
\hspace{-0.5cm}&&\Sigma_{\rm ep}(i\omega_m,{\bf k})=-\frac{1}{\beta}\sum_{n}\int \frac{d{\bf k}'}{(2\pi)^2}\int
\frac{dq_z}{2\pi}|g_{{\bf k}',q_z}|^2\\
\hspace{-0.5cm}&&\mbox{}\times\frac{2\omega_{{\bf
      k}',q_z}}{(i\omega_{n})^2-\omega^2_{{\bf k}',q_z}}
G(i\omega_m-i\omega_{n},{\bf k}-{\bf k}').
\label{ep_finite}
\end{eqnarray}
In Eq.~(\ref{ep_finite}), $g_{{\bf k}',q_z}$ denote the e-p
interactions due to the deformation potential (LA-branch) and piezoelectric
coupling (LA and TA branches), and $\omega_{{\bf k},q_z}$ are the
corresponding energy spectra. For LA and TA phonons, $\omega^{sl}_{{\bf
    k},q_z}=v_{sl}\sqrt{k^2+q_z^2}$ and $\omega^{st}_{{\bf
    k},q_z}=v_{st}\sqrt{k^2+q_z^2}$, respectively, with $v_{sl}$ and $v_{st}$
being the velocities of LA and TA phonons, respectively.\cite{e_p_formula1,e_p_formula2}

From the structure of the self-energies due to the e-e and e-p interactions,
  one observes that every elements in these $4\times 4$ matrices are renormalized,
  including the effective mass, the zero-energy point, the strength of
the SOC and the singlet order parameter. Specifically, the triplet
order parameter is induced due to the existence of the triplet pairings.
 Here, we neglect the renormalization of the
effective mass, the zero-energy point and the SOC strength (which is shown to be
negligible compared to the original SOC), and focus on the renormalization of
the singlet and induction of the triplet order
parameters, whose concrete analytical expressions and numerical values are discussed
 in detail in Sec.~\ref{numerical}.

\section{Numerical Results}
\label{numerical}
In this section, to show the physics more clearly and quantitatively, 
we numerically calculate the self-energies due to the e-e and e-p interactions
based on Eqs.~(\ref{ee_finite}) and (\ref{ep_finite}). We
choose the material with strong SOC: i.e., InSb (100) QWs. 
All parameters including the band structure
and material parameters used in our computation are
listed in Table~\ref{material}.\cite{material_book,parameter_SOC}

\begin{table}[h]
\caption{Parameters used in the computation for self-energies due to the e-e and e-p
  interactions.\cite{material_book,parameter_SOC}}
 \label{material} 
\begin{tabular}{ll|ll}
    \hline
    \hline    
    $m^*/m_0$&\;\;\;$0.015$&\;$n_0$~(cm$^{-2}$)&\;\;\;$10^{14}$\\[4pt]
    $\kappa_0$&\;\;\;$16.0$ &\;$\gamma_D~({\rm eV}\cdot{\rm \mathring{A}}^3)$&\;\;\;$389$\\[4pt]
    $\kappa_{\infty}$&\;\;\;$15.68$&\;$a~({\rm nm})$&\;\;\;$3$\\[4pt]
    $d~({\rm kg/cm^3})$&\;\;\;$5.8$&\;$T~({\rm K})$&\;\;\;$2$ \\[4pt]
    $\Xi~({\rm eV})$&\;\;\;$14.5$ &\;$v_{sl}~({\rm m/s})$&\;\;\;$3770$ \\[4pt]
    $e_{\rm 14}~({\rm 10^9~V/m})$&\;\;\;$1.41$ &\;$v_{st}~({\rm m/s})$&\;\;\;$1630$\\[4pt]
    \hline
    \hline
\end{tabular}
\end{table}
In Table~\ref{material}, $d$ is the mass density of the crystal; $\Xi$ denotes the
  deformation potential; and $e_{14}$ represents the piezoelectric constant.
 In our computation, the electron densities $n_e$ in QWs 
vary from $n_0$ to $35n_0$. With these electron
densities, the chemical potential is calculated with the strong SOC explicitly
included in the energy spectra by solving the equation  
\begin{equation}
n_{\uparrow}=n_{\downarrow}=\frac{1}{2}\int
\frac{d{\bf k}}{(2\pi)^2}\big[n_F(\epsilon_{{\bf k}+})+n_F(\epsilon_{{\bf k}-})\big].
\label{chemical}
\end{equation}
In Eq.~(\ref{chemical}), $n_{\uparrow}$ and $n_{\downarrow}$ represent the electron densities with
spin-up and spin-down, respectively; $n_F(\epsilon_{{\bf k}\pm})=\Big\{\exp\big[\beta(\epsilon_{{\bf
    k}\pm}-\mu)\big]+1\Big\}^{-1}$ is the Fermi-Dirac distribution function.
Furthermore, in our computation, we focus on the
 {\it weak coupling limit} addressed in Refs.~\onlinecite{Dar_Sarma1,Dar_Sarma2,Hubbard1} with
  $|\Delta(\omega,{\bf k})|\ll \Delta_0$, where $\Delta_0$ is one to several
  meV in conventional superconductors. Moreover, we focus on the low temperature
limit. With these two conditions, one observes that the main physics happens in
the regime $|\omega|\lesssim |\Delta(\omega,{\bf k})|$, and hence the
frequency is much smaller
than $\Delta_0$. In this situation, in the singlet order parameter due to the
superconducting proximity effect, i.e., $\Delta(\omega,{\bf k})$
  [Eq.~(\ref{effective_potential})], the frequency dependence can be
  neglected.\cite{Dar_Sarma1,Dar_Sarma2,Hubbard1}
Therefore, in our calculation, $\Delta(\omega,{\bf k})$ is set to be constant (0.5~meV) in
  the static approximation. 
 It is emphasized that this approximation has little qualitative influence
on the physics we reveal.\cite{Dar_Sarma1,Dar_Sarma2,Hubbard1}

Finally, we point out that according to our
calculation based on above parameters, it is found that the contribution of the self-energy mainly comes from
the e-e interaction, as the contribution due to e-p interactions are
two orders of magnitude
smaller than that of the e-e interaction at low temperature. Accordingly, it is adequate to consider
the e-e interaction here in the calculation and the following analysis.

\subsection{Suppression of singlet order parameter}
\label{SS}

In this part, we focus on the calculation of the Coulomb-interaction--induced singlet order
  parameter $\Delta_{s}({\bf k})$.
From Eqs.~(\ref{FF}) and (\ref{ee_finite}), the Coulomb-interaction--induced singlet order
  parameter
 in the static approximation is obtained, which is written as 
\begin{equation}
\Delta_{s}({\bf k})\approx-\frac{\Delta}{2\beta}\sum_{{\bf k}',n}\sum_{\eta=\pm}V_{{\bf k}-{\bf
    k}'}\frac{1}{\omega_n^2+\epsilon^2_{{\bf
    k}'\eta}+|\Delta|^2}.
\label{opposite}
\end{equation}
One notes that according to Eq.~(\ref{effective}), the
Coulomb-interaction--induced 
order parameter is defined from the self-energy multiplying $\hat{\tau}_3$.
From Eq.~(\ref{opposite}), one observes that $\Delta_{s}({\bf k})$ always has
opposite sign against the proximity-induced order parameter
 $\Delta(\omega=0)$, because the summation in Eq.~(\ref{opposite}) is
always positive with the repulsive e-e Coulomb interaction.
 This shows that the repulsive
Coulomb interaction suppresses the singlet order parameter
with the renormalized singlet order parameter 
$\Delta_S({\bf k})=\Delta+\Delta_{s}({\bf k})$.
 It is further noted that this conclusion is consistent with the recent 
investigation in quantum nanowire in proximity to the $s$-wave superconductor,
in which the Hubbard interaction is considered.\cite{Hubbard2}

Furthermore, after the summation on the Matsubara frequencies, Eq.~(\ref{opposite})
becomes \begin{eqnarray}
\nonumber
\hspace{-0.3cm}&&\Delta_{s}({\bf
  k})=-\frac{m^*}{16\pi^2}\sum_{\eta=\pm}\int_0^{\infty}d\varepsilon_{\bf
  k'}d\phi_{\bf {k'}}F_{{\bf k},{\bf
    k'}}\Lambda_{\eta}({\bf k'})\\
\hspace{-0.3cm}&&\mbox{}\times\Big[1-2n_F\big(\sqrt{\epsilon_{{\bf
      k'},\eta}^2+|\Delta|^2}\big)\Big],
\label{complicated}
\end{eqnarray}
with $F_{{\bf k},{\bf k}'}=V\big(\sqrt{k^2+k'^2-2kk'\cos\phi_{\bf k'}}\big)$, 
which is $\phi_{\bf k}$-independent, and $\Lambda_{\pm}({\bf
  k'})=\Delta/\sqrt{\epsilon_{{\bf
      k'},\pm}^2+|\Delta|^2}$.
Accordingly, from Eq.~(\ref{complicated}), one observes that the renormalized
  singlet order parameter only depends on the magnitude of the momentum and is
  independent on its direction, which is calculated explicitly in the following.

\subsubsection{Momentum dependence of the renormalized singlet order parameter}

\begin{figure}[ht]
  {\includegraphics[width=8.5cm]{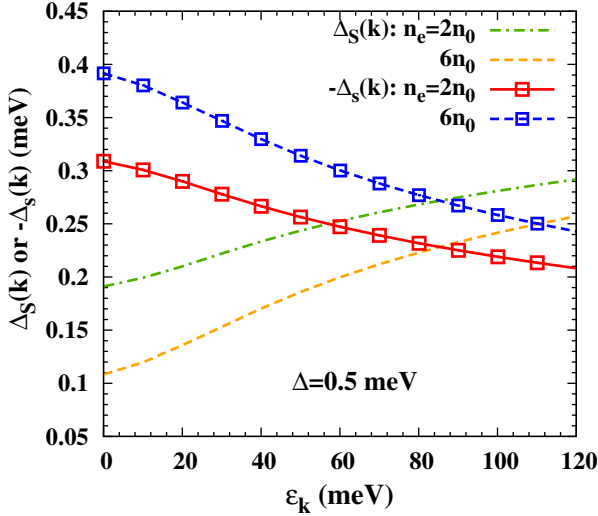}}
  \caption{(Color online) Energy-dependencies of the renormalized
  singlet order parameter $\Delta_{S}({\bf k})$ and the magnitude of the 
  Coulomb-interaction--induced singlet
  order parameter $-\Delta_{s}({\bf k})$ 
with different electron densities $n=2n_0$ and $6n_0$, respectively.}
  \label{figyw2}
\end{figure}

In this part, we study the momentum dependence of the
renormalized singlet
  order parameter, which only depends on the magnitude of the momentum.
 In Fig.~\ref{figyw2}, the renormalized singlet order parameters
    $\Delta_{S}({\bf k})$, shown by the green 
    chain and yellow dashed curves
    for $n_e=2n_0$ and $6n_0$, increase with the
increase of the electron energy. This is because the magnitude of the  
  Coulomb-interaction--induced singlet
  order parameters, i.e., $-\Delta_{s}({\bf k})$, decreases with the
  increase of the electron energy for 
   $n=2n_0$ (the red solid curve with squares) and
  $6n_0$ (the blue dashed curve with squares),
 respectively. This can be undersrood from the fact that with the
  increase of the magnitude of the momentum and hence the electron energy, the
    Coulomb interaction is suppressed.
 
Furthermore, in Fig.~\ref{figyw2}, by observing the calculated results with $n_e=2n_0$ and
$6n_0$, one notices that the Coulomb-interaction--induced singlet order
  parameter and hence the
renormalized singlet order parameter explicitly depends on the electron density in
QWs. Actually, this provides a possible way to experimentally distinguish the singlet order parameter
due to the superconducting proximity
 effect and that due to the e-e interaction. This is because the singlet order parameter due to the
superconducting proximity effect marginally depends on the electron density in QWs
[Eq.~(\ref{effective_potential})].

\subsubsection{Electron density dependence of the 
  Coulomb-interaction--induced singlet order parameter}
\label{singlet_density}

\begin{figure}[ht]
  {\includegraphics[width=8.5cm]{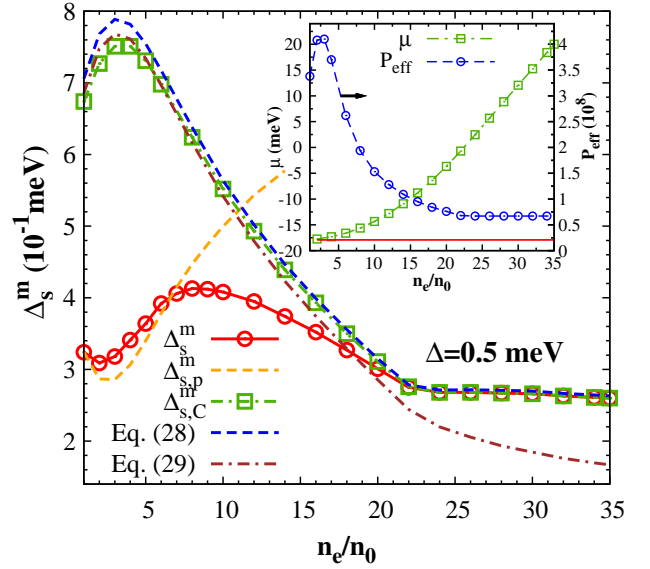}}
  \caption{(Color online) Density dependence of the maximum of the
    Coulomb-interaction--induced singlet order parameter
  $\Delta_{s}^m$, shown by the red solid curve with circles. The yellow dashed
  curve (green
  chain curve with squares), labeled by $\Delta_{s,p}^m$ 
  ($\Delta_{s,C}^m$), represents the
  maximum of the Coulomb-interaction--induced singlet order parameter when the proximity-induced
  singlet pairing 
  (Coulomb potential) is arbitrarily taken to be independent on electron
  density. Furthermore, the blue dashed and purple
chain curves denote the calculated results from Eqs.~(\ref{simplified}) and
(\ref{simplified_further}), respectively, in which the
longitudinal polarization function is also taken to be constant.
The inset zooms the density dependencies of the chemical potential and
  effective polarization function [Eq.~(\ref{effective_screening})], in which the
  red solid line represents the band edge.}
  \label{figyw3}
\end{figure}

In this part, we focus on the electron density
  dependence of the maximum of the Coulomb-interaction--induced singlet order parameter
  $\Delta_{s}^m$ at ${\bf k}=0$.
 In Fig.~\ref{figyw3}, it is shown by the red solid curve with circles
 that with the increase of the electron
  density, $\Delta^m_{s}$ first shows a valley at
  relatively low electron density $n_e\approx 3n_0$, then a peak at the
  moderate electron density $n_e\approx 8n_0$, and finally decreases very slowly
 at high electron density $n_e\gtrsim 24n_0$. 
We first give the whole physics picture behind these
 rich and intriguing dependencies of $\Delta^m_{s}$ from the
 analysis of 
 Eq.~(\ref{opposite}) when ${\bf k}=0$.

From Eq.~(\ref{opposite}), one finds that with the increase of the electron
density, both the Coulomb potential ($V_{\bf k'}$) and the proximity-induced
  singlet pairing
 ($\frac{\displaystyle 1}{\displaystyle \omega_n^2+\epsilon^2_{{\bf
    k}'\eta}+|\Delta|^2}$) are varied due to their dependencies on the chemical
potential, and hence both can influence of the
electron density dependence of $\Delta^m_{s}$. Specifically,
 in the inset of Fig.~\ref{figyw3}, it is shown by the blue dashed curve with
 circles that with the increase
  of the electron density, the effective polarization function 
\begin{equation}
 P_{\rm eff}=e^2/(\varepsilon_0\kappa_0)|P^{(1)}(\omega=0,{\bf
    q}=0)|.
\label{effective_screening}
\end{equation} 
shows a peak arising at the
  crossover between the
  non-degenerate and degenerate regimes with $n_e\approx 3n_0$, and becomes independent of the electron
  density when $n_e\gtrsim 20n_0$. Therefore,  
the strength of the Coulomb potential first decreases at low electron density
and then increases at moderate one, and finally becomes independent of
electron density at high electron density. 
As to the proximity-induced
  singlet pairing, with the increase
  of the electron density, it first varies slowly and then rapidly due to the
  electron density dependence of the chemical potential, which is 
 shown by the green chain curve with squares in the inset of Fig.~\ref{figyw3}.

We find that the valley (decrease) in the electron density dependence of
 $\Delta_{s}^m$ at low (high) electron density comes from 
the electron density dependence of the Coulomb potential (proximity-induced
  singlet pairing). This is confirmed from the fact that when the electron density is low (high),
 $\Delta_{s,p}^m$ ($\Delta_{s,C}^m$) with constant proximity-induced
  singlet pairing
 (constant Coulomb potential) at $n_e=n_0$ ($n_e=35n_0$) almost coincides
 with $\Delta_{s}^m$, shown by the yellow dashed curve (green chain
 curve with squares) in Fig.~\ref{figyw3}.
 Accordingly, at the moderate
  electron density $3n_0\lesssim n_e\lesssim 20n_0$, with the increase of the
  electron density, the Coulomb potential tends to
enhance $\Delta_{s}^m$; whereas the proximity-induced
  singlet pairing tends to suppress
$\Delta_{s}^m$. Thus, due to 
this competition of these two effects, a peak arises at the moderate
electron density. 

 Nevertheless, 
one observes that for $\Delta_{s}^m$ and $\Delta_{s,p}^m$, there
  also exists small discrepancy for the value of the valley, with the former
  larger than the latter. This can be
  explained by the fact that when the
  Coulomb potential is set to be density independent,
 $\Delta_{s,C}^m$ shows a peak when $n_e\approx 3n_e$, which
  suppresses the value of valley. Moreover,
  although when $n_e\gtrsim 8n_e$, $\Delta_{s}^m$ decreases with the increase
  of the electron density, the rates of the decrease are drastically different
  for $8n_0\lesssim n_e\lesssim 20n_e$ and $n_e\gtrsim 20n_e$.
 This can also be explained by the electron density dependence of
 $\Delta_{s,C}^m$, which decreases rapidly when $8n_0\lesssim n_e\lesssim
  20n_e$ and slowly when $n_e\gtrsim 20n_e$.    
 Accordingly, a detailed physics picture can be
  obtained by analyzing the electron density dependence of
  $\Delta_{s,C}^m$, which is addressed as follows.

\begin{figure}[ht]
  {\includegraphics[width=7cm]{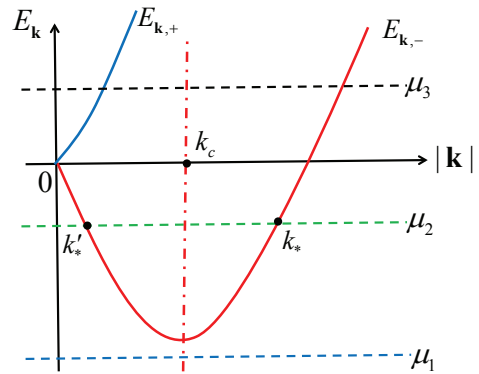}}
  \caption{(Color online) Schematic for
 the band structures of $E_{{\bf k},-}$ and $E_{{\bf
          k},+}$-bands, shown by the red and blue solid curves, respectively.
 The dashed lines labeled by $\mu_1$,
  $\mu_2$ and $\mu_3$ correspond to the chemical potentials when $n_e\lesssim
  3n_0$, $3n_0\lesssim n_e\lesssim 20n_0$ and $n_e\gtrsim 20n_0$,
  respectively. $k_c$ is the momentum corresponding to the band edge of $E_{{\bf
      k},-}$-band; $k_*$ and $k_*'$ label the intersection points between
  $\mu_2$ and $E_{{\bf
      k},-}$-band with $k_*>k_c$ and $k'_*<k_c$, respectively.}
  \label{figyw4}
\end{figure}

 \paragraph{Single-band and double-band regimes}
To analyze the electron density dependence of
  $\Delta_{s,C}^m$, we first divide the system into different regimes
  according to the population of electrons with different chemical potentials.
 In the inset
  of Fig.~\ref{figyw3},  when $n_e\lesssim 3n_0$, the chemical potential is
  close to the band edge,
 which is shown by the
  red solid line, indicating that the
  system lies in the crossover between the non-degenerate and degenerate regimes. Moreover, when
  $n_e\lesssim 20n_0$, the chemical potential is negative, which shows that the
  electrons mainly populate at the $E_{{\bf k},-}$-band; whereas when
  $n_e\gtrsim 20n_0$, $E_{{\bf k},+}$-band becomes populated. 
To see this more clearly, in Fig.~\ref{figyw4}, the band structures for the $E_{{\bf k},-}$ and $E_{{\bf
      k},+}$-bands are schematically plotted by the red and blue solid curves,
  respectively.
 In Fig.~\ref{figyw4}, 
 the three situations mentioned above corresponding to $n_e\lesssim
  3n_0$, $3n_0\lesssim n_e\lesssim 20n_0$ and $n_e\gtrsim 20n_0$ are plotted by
  the dashed lines labeled by $\mu_1$,
  $\mu_2$ and $\mu_3$, respectively.  
Accordingly, when $n_e\lesssim 20n_0$, only $E_{{\bf k},-}$-band is
  efficiently populated, the system is referred to as {\it single-band
  regime}; whereas when
$n_e\gtrsim 20n_0$, the $E_{{\bf k},+}$-band becomes populated and the system 
is referred to as {\it double-band regime}.

\paragraph{Influence of proximity-induced singlet pairing in different regimes}
Before addressing the influence of the proximity-induced singlet pairing in
different regimes, we first make some simplification in Eq.~(\ref{complicated}). One
 notes that when $\Delta\gg k_BT$ here, $n_F\big(\sqrt{\epsilon_{{\bf
      k'},\pm}^2+|\Delta|^2}\big)\ll 1$ and hence can be neglected in
Eq.~(\ref{complicated}). This is justified in Fig.~\ref{figyw3} by the fact that
when $n_F\big(\sqrt{\epsilon_{{\bf
      k'},\pm}^2+|\Delta|^2}\big)$ is not considered in Eq.~(\ref{complicated}),
$\Delta_{s,C}^m$ shown by
the blue dashed curve almost
coincides with the green chain
curve with squares. In this situation, Eq.~(\ref{complicated}) with ${\bf k}=0$ is simplified to
be 
\begin{equation}
\Delta^m_{s,C}\approx\frac{m^*}{8\pi}\int_0^{\infty}d\varepsilon_{\bf k'}F^*_{{\bf
    k'}}\big[\Lambda_{-}({\bf k'})+\Lambda_{+}({\bf k'})\big],
\label{simplified}
\end{equation}
where $F^*_{{\bf k'}}=V_{{\bf k}=0,{\bf k'}}$ with polarization function taken to
be the one when $n_e=35n_0$.
 Nevertheless, one further notes that when $n_e\lesssim 20n_0$, only the 
$E_{{\bf k},-}$-band is efficiently populated and hence $|\epsilon_{{\bf k},-}|\ll
|\epsilon_{{\bf k},+}|$. Accordingly, Eq.~(\ref{simplified}) is further reduced to 
\begin{equation}
\Delta^m_{s,C}\approx\frac{m^*}{8\pi}\int_0^{\infty}d\varepsilon_{\bf k'}F^*_{{\bf
    k'}}\Lambda_{-}({\bf k'}).
\label{simplified_further}
\end{equation}
  Eq.~(\ref{simplified_further}) is justified in Fig.~\ref{figyw3} with
the fact that when $n_e\lesssim 20n_0$,
$\Delta_{s,C}^m$ shown by the purple chain curve almost
coincides with the blue dashed one.

When the system lies in the
single-band (double-band) regime when $n_e\lesssim 20n_0$ ($n_e\gtrsim 20n_0$),
 the electron density dependence of $\Delta_{s,C}^m$ can be analyzed based on
Eq.~(\ref{simplified_further}) [Eq.~(\ref{simplified})]. Specifically, 
in the single-band regime, when $n_e\lesssim 3n_0$,
 the kinetic energy of electrons is larger than the
chemical potential (refer to the inset in Fig.~\ref{figyw3}). 
In this situation, with the increase of the electron density and hence the chemical
potential, $\epsilon_{{\bf
      k'},-}^2$ decreases, leading to the increase of  $\Delta^m_{s,C}$
 from Eq.~(\ref{simplified_further}). 
Whereas when
 $3n_0\lesssim n_e\lesssim 20n_0$, the chemical potential is larger 
 than the band edge. This situation is represented in Fig.~\ref{figyw4} with the
chemical potential $\mu_2$ intersecting with the $E_{{\bf
      k},-}$-band by two points $k_*$ and $k_*'$. Specifically, $k_*>k_c$ and
  $k'_*<k_c$ with $k_c$ being the momentum corresponding to the band edge of $E_{{\bf
      k},-}$-band. From Eq.~(\ref{simplified_further}), one observes that
 when $\epsilon_{{\bf k},-}=0$, $\Lambda_{-}({\bf k})$
is the largest, which means that the electrons around the chemical potential  
play the most important role in the renormalization of the singlet order
parameter.  For these electrons, with the increase of the chemical potential
  $\mu_2$, $k_*$ increases and $k_*'$ decreases. Nevertheless, $k_*'$ is
  relatively small and can be even smaller than the wave-vector due to the
  effective polarization function, which cannot cause efficient variation
  of the Coulomb potential. Whereas the increase of $k_*$ can efficiently
  suppress the Coulomb potential, and hence $\Delta^m_{s,C}$ decreases with
  the increase of the electron density in this regime. Furthermore, when $n_e\gtrsim 20n_0$,
  the system enters into the double-band regime, in which the $E_{{\bf
      k},+}$-band becomes populated. Therefore, with the increase of the population of these electrons,
  the contribution of the $E_{{\bf
      k},+}$-band to $\Delta^m_{s,C}$ increases with the increase of
  the electron density [Eq.~(\ref{simplified})]. This tends to suppress the
  decrease of $\Delta_{s,a}^m$ due to the $E_{{\bf
      k},-}$-band. Consequently, in this regime, $\Delta_{s,C}^m$ decreases very
  slowly with the increase of the electron density.  
  
\paragraph{Summary of the physics picture}

By knowing the separate roles of the Coulomb potential and proximity-induced
singlet pairing in the
  density dependence of $\Delta_{s}^m$, the whole physics picture can be
  obtained. At the low
  electron density $n_e\lesssim 3n_0$, from the inset of Fig.~\ref{figyw3}, one
  observes that the polarization function varies rapidly, whereas the
  chemical potential varies slowly. In this situation, the influence of the
  Coulomb potential on the renormalized singlet order parameter is
  dominant. As a consequence, at the crossover of the
  non-degenerate and degenerate regimes with $n_e\approx 3n_0$, there exists a valley in the electron
  density dependence of $\Delta_{s}^m$ due to the dependence
  of the screening effect on electron density. Nevertheless, at the moderate
  electron density $3n_0\lesssim n_e\lesssim 20n_0$, the screening effect tends to
enhance $\Delta_{s}^m$; whereas the population of the electron in
$E_{{\bf k},-}$-band tends to suppress it. Thus, due to 
this competition of the two effects, a peak arises in the electron
  density dependence of the $\Delta_{s}^m$. Finally, at the high
  electron density $n_e\gtrsim 20n_0$, from the inset of Fig.~\ref{figyw3}, one
  obtains that the effective polarization function becomes constant, whereas the
  chemical potential increases rapidly. Therefore, $\Delta_{s}$ decreases slowly
  due to the competition of the populations of electrons in $E_{{\bf k},+}$ and
  $E_{{\bf k},-}$-bands.

\subsection{Induced triplet $p$-wave order parameter}
\label{PP}
In this subsection, we discuss the triplet order parameter
  induced by the e-e Coulomb interaction.
 Specifically, from Eqs.~(\ref{FF}) and (\ref{ee_finite}), 
the induced triplet order parameter reads  
\begin{equation}
\Delta_{t}({\bf k})=
\left(\begin{array}{cc}
Q_{-}({\bf k})-Q_{+}({\bf k}) & 0\\
0 &Q^*_{+}({\bf k})-Q^*_{-}({\bf k})
\end{array}\right).
\label{triplet2}
\end{equation}
with
\begin{eqnarray}
\nonumber
&&Q_{\pm}({\bf k})=\frac{m^*}{8\pi^2}e^{i\phi_{\bf
    k}}\int d\varepsilon_{{\bf k}'}d\phi_{{\bf k}'}\cos\phi_{\bf
    k'}F_{{\bf k},{\bf k}'}\Lambda_{\pm}({\bf k'})\\
&&\times\Big[\frac{1}{2}
-n_F(\sqrt{\epsilon_{{\bf
      k}',\pm}^2+\Delta^2})\Big].
\end{eqnarray}
 From Eq.~(\ref{triplet2}), one obtains that the triplet order parameter
 $\Delta_{t}({\bf k})$ depends on the phase factor $e^{i\phi_{\bf k}}$
 and hence the direction of the momentum, which is odd in the momentum. Thus,
$\Delta_{t}({\bf k})$
 is the triplet $p$-wave order parameter. Specifically, this
   $p$-wave order parameter is in the ($p_x\pm ip_y$)
   type.\cite{Sigrist,Sigrist_PRL,SrRuO_triplet,Leggett}
 It is noted that if $F_{{\bf k},{\bf
       k}'}$ is arbitrarily taken to be momentum-independent,
   the triplet order parameter [Eq.~(\ref{triplet2})] is exactly zero due to the angle integration.

\subsubsection{Momentum dependence of triplet $p$-wave order parameter}
\label{triplet_momentum}
\begin{figure}[ht]
  {\includegraphics[width=8cm]{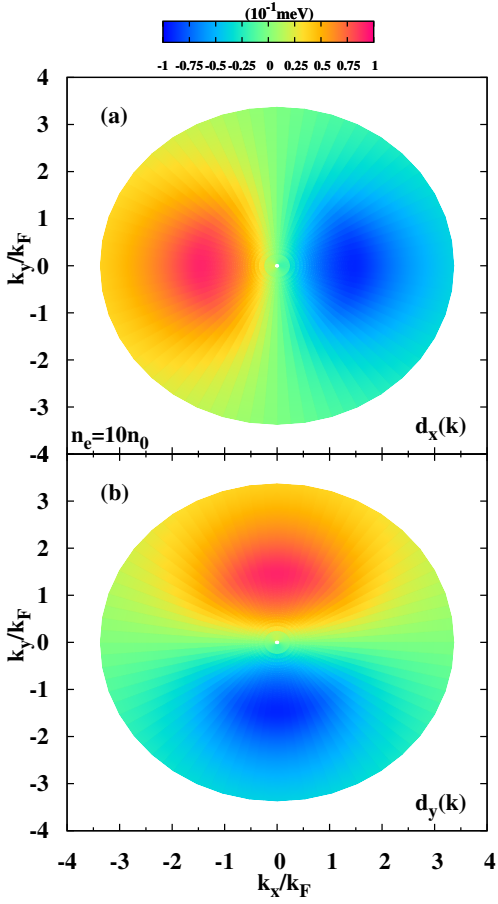}}
  \caption{(Color online) Momentum dependence of the ${\bf
      d}$-vectors. The electron density is $n_e=10n_0$ with $k_F=7.9\times 10^7$~m$^{-1}$. In (a)
    and (b), it is shown that the $\hat{\bf x}$ and $\hat{\bf y}$ components of the induced triplet order
  parameter satisfy $d_x({\bf k})\propto -\cos\phi_{\bf k}$ and $d_x({\bf
    k})\propto \sin\phi_{\bf k}$, respectively. Moreover, the calculated
results show that the induced triplet order parameter also
depends on the magnitude of momentum.}
  \label{figyw5}
\end{figure}

In this part, we analyze the momentum dependence of the triplet order parameter
  including the angular and magnitude dependencies, which is summarized in
  Fig.~\ref{figyw5}.
In Fig.~\ref{figyw5}, when $n_e=10n_0$, the ${\bf d}$-vectors of the triplet
order parameter are plotted, defined as\cite{Triplet_S_F,Sigrist,Sigrist_book} 
\begin{equation}
\Delta_{t}({\bf k})=\big[{\bf d}({\bf k})\cdot{\bgreek \sigma}\big]i\sigma_y,
\end{equation}
are plotted, with ${\bf d}({\bf k})=\big(d_x({\bf k}),d_y({\bf k}),d_z({\bf k})\big)$.
It is shown that only the in-plane components
of the ${\bf d}$-vector are induced in (100) QWs. Specifically, in Figs.~\ref{figyw5}(a)
    and (b), it is shown that the $\hat{\bf x}$ and $\hat{\bf y}$ components of the induced triplet order
  parameter satisfies $d_x({\bf k})\propto -\cos\phi_{\bf k}$ and $d_y({\bf
    k})\propto \sin\phi_{\bf k}$, which is parallel to the effective
  magnetic field due to the SOC [Eq.~(\ref{SOC})]. Thus, the induced triplet
  order parameter is stable due to the SOC.\cite{Sigrist_PRL} 
 Moreover, the calculated
results in Fig.~\ref{figyw5} show that the induced triplet order parameter also
depends on the magnitude of momentum or electron energy, which are further
discussed in the following.

\begin{figure}[ht]
  {\includegraphics[width=8.5cm]{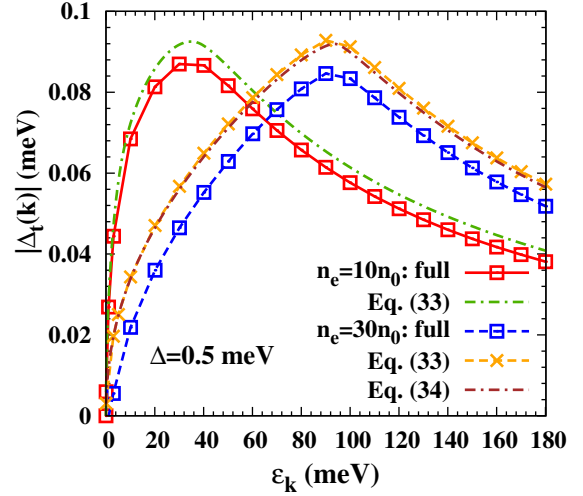}}
  \caption{(Color online) Energy dependence of the absolute value of the induced
    triplet order parameter $|\Delta_{t}({\bf k})|$ when $n_e=10n_0$
    (the red solid curve with squares) and $30n_0$ (the blue dashed curve with
    squares). Furthermore, the green chain curve (yellow dashed curve with crosses) represents the calculated
    results based on Eq.~(\ref{triplet_simplified})
     when $n_e=10n_0$ ($n_e=30n_0$). Moreover, from
     Eq.~(\ref{triplet_simplified2}), the calculated result for $n_e=30n_0$ is
     plotted by the purple chain curve.}
  \label{figyw6}
\end{figure}  

In Fig.~\ref{figyw6}, the energy dependence of the 
absolute value of the induced triplet order parameters $|\Delta_{t}({\bf k})|$
 is plotted when $n_e=10n_0$ and $30n_0$, respectively. 
It is shown that with the increase of the
electron energy, $|\Delta_{t}({\bf k})|$ first increases from zero
at ${\bf k}=0$ and then decreases with a peak arising at moderate energy. This can be understood as
follows. When ${\bf k}=0$, from Eq.~(\ref{triplet2}), $\Delta_{t}({\bf
  k})$ equals to zero due to the angular integration over $\phi_{\bf k'}$. To
further analyze the energy dependence of $|\Delta_{t}({\bf k})|$,
Eq.~(\ref{triplet2}) can be simplified, which is similar to the situation of
the Coulomb-interaction--induced singlet
order parameter.
 It can be seen that $n_F(\sqrt{\epsilon_{{\bf
      k}',\pm}^2+\Delta^2})$ can be neglected in Eq.~(\ref{triplet2}) when
$\Delta\gg k_BT$. Furthermore, when $n_e\lesssim 20n_0$, the system lies in
the singlet-band regime, and the magnitude of the triplet order parameter is
written as
\begin{equation}
|\Delta_{t}({\bf k})|\approx\frac{m^*}{16\pi^2}
\int d\varepsilon_{{\bf k}'}d\phi_{{\bf k}'}
F_{{\bf k},{\bf k}'}\cos(\phi_{\bf k'})\Lambda_{-}({\bf k'}).
\label{triplet_simplified}
\end{equation}
Whereas when $n_e\gtrsim 20n_0$, the system lies in the double-band regime, with
the corresponding magnitude of the triplet order parameter written as
\begin{eqnarray}
\nonumber
&&|\Delta_{t}({\bf k})|\approx\frac{m^*}{16\pi^2}
\int d\varepsilon_{{\bf k}'}d\phi_{{\bf k}'}
F_{{\bf k},{\bf k}'}\cos(\phi_{\bf k'})\\
&&\mbox{}\times\big[\Lambda_{-}({\bf k'})-\Lambda_{+}({\bf k'})\big].
\label{triplet_simplified2}
\end{eqnarray}
Based on Eqs.~(\ref{triplet_simplified}) and (\ref{triplet_simplified2}),
 the energy dependencies of
$|\Delta_{t}({\bf k})|$ can be understood as follows.

We first address that when ${\bf k}$ is very large, the Coulomb
  potential is efficiently suppressed, and hence the induced triplet order
  parameter tends to be zero when the energy tends to infinite. Accordingly, when
  $|{\bf k}|=0$ and $|{\bf k}|\rightarrow \infty$, $|\Delta_{t}({\bf
    k})|\rightarrow 0$. Therefore, there must exist non-monotonic behavior
  between $|{\bf k}|=0$ and $|{\bf k}|\rightarrow \infty$, which is shown to be a
  peak at the moderate energy. 
Specifically, when $n=10n_0$, the system lies in the single-band regime and the behavior
  of $|\Delta_{t}({\bf k})|$ can be described by
  Eq.~(\ref{triplet_simplified}) well. This is justified in Fig.~\ref{figyw6}
 by the fact that the green chain curve calculated from Eq.~(\ref{triplet_simplified}) almost
coincides with the full calculation represented by the red solid curve with
squares. From
Eq.~(\ref{triplet_simplified}), on one hand, one observes that when $|{\bf k}'|\approx |{\bf
k}|$, the Coulomb potential is relatively strong. Accordingly, electrons with
momentum $|{\bf k}'|\approx |{\bf
k}|$ can play an important role in the induction of the triplet order
parameter. On the other hand, when $\epsilon_{{\bf k},-}=0$, $\Lambda_{-}({\bf k})$
is largest, which means the electron around the chemical potential can also
play an important role in the induction of the triplet order
parameter. When the two parts of the electrons, i.e., the electrons with momentum
$|{\bf k}|$ and the ones around the chemical potential, are not the same, the
induced triplet order parameter is expected to be small. However, with the increase of the
momentum ${\bf k}$, there exists a ``intersection'' point that the two parts of the
electrons are the same, where the peak arises in the energy dependence of
$|\Delta_{t}({\bf k})|$. Accordingly, from this simple picture, the position
of the peak of $|\Delta_{t}({\bf k})|$ in the energy dependence can be determined.

 Specifically, one expects that in the single-band
regime, the ``intersection'' point arises when the condition $k_c^2/(2m^*)-\alpha
k_c-\mu\approx 0$ is satisfied, where $k_c$ is the magnitude of the momentum at
the ``intersection''
point. Therefore, the ``intersection'' point is estimated to be
\begin{equation}
\varepsilon^c_k\equiv k_c^2/(2m^*)\approx
m^*\alpha^2+\mu+\alpha\sqrt{2m^*(m^*\alpha^2/2+\mu)},
\label{magic_position}
\end{equation}
which labels the position of the peak. Here, we compare
Eq.~(\ref{magic_position}) with the full numerical calculations. From
Eq.~(\ref{magic_position}), 
 when $n_e=5n_0$, $10n_0$ and $20n_0$, the calculated peak position are
 at 27.0, 37.4 and 64.1~meV, respectively. They are very close to the corresponding ones from the full
 numerical calculations, which are 27.3, 35.1 and 63.4 meV. Furthermore, from
 Eq.~(\ref{magic_position}), one obtains that when the electron density and
 hence the chemical potential increases, the position of the peak arises at
 higher energy.

When $n_e=30n_0$, one expects that the system enters into the double-band
  regime. However, it is shown in Fig.~\ref{figyw6} that the results calculated from
  Eqs.~(\ref{triplet_simplified2}) and (\ref{triplet_simplified}) almost
  coincide, denoted by the purple chain curve and yellow dashed curve with
  crosses, respectively. This indicates that the contribution from $E_{{\bf
      k},+}$-band is negligible even when its population becomes significant. This is
because even in the double-band regime, the average momentum of the populated electrons in $E_{{\bf
      k},+}$-band is close to zero, which is much smaller than the ones in $E_{{\bf
      k},-}$-band. This can be explicitly seen from Fig.~\ref{figyw4} that
    the momentum corresponding to the intersection point between $\mu_3$ and $E_{{\bf
      k},+}$-band is close to zero. Specifically, the average momentum in $E_{{\bf
      k},+}$-band is much smaller than the wave-vector due to the
    effective polarization function. Hence,
  the Coulomb potential experienced by the 
  electron in $E_{{\bf
      k},+}$-band can be treated as momentum-independent potential approximately,
  which does not contribute to the induction of the triplet order parameter due
  to the angle integration in Eq.~(\ref{triplet2}).\cite{Gorkov_Rashba} In this
  situation, the system still lies in the single-band regime. Hence, the
  position of the peak is determined from Eq.~(\ref{magic_position}), which
is calculated to be 94.2~meV, again very close to the one from the full
calculation, i.e., 93.7~meV.

\subsubsection{Electron density dependence of triplet order parameter}

\begin{figure}[ht]
  {\includegraphics[width=8.4cm]{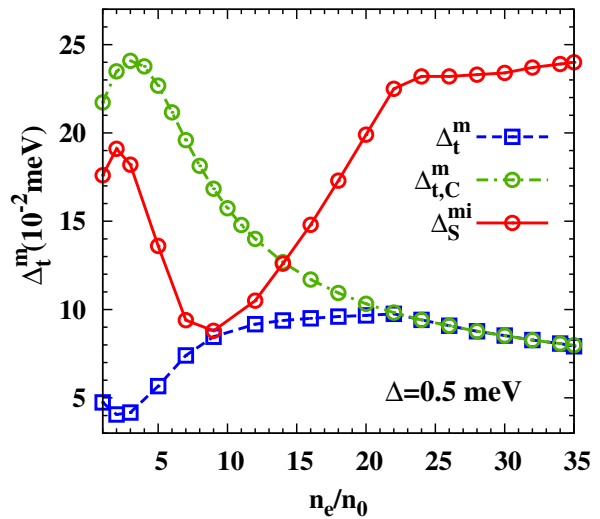}}
  \caption{(Color online) Electron density dependence of the maximum value of
    the induced triplet order parameter $\Delta_{t}^m$, shown by the blue
    dashed curve with squares. The green chain curve with circles denoted
    by $\Delta_{t,C}^{m}$
    represents the calculated results when the polarization function is taken to
    the one when $n_e=35n_0$
in Eq.~(\ref{triplet2}). Finally, the red solid curve with circles represents the
minimum value of the renormalized singlet order parameter $\Delta_{S}^{mi}$.}
  \label{figyw7}
\end{figure}

In this part, we study the electron density dependence of the maximum value of the triplet order
parameter $\Delta_{t}^m$, shown in Fig.~\ref{figyw7}.  
It is seen that with
the increase of the electron density, $\Delta_{t}^m$ first decreases at the
low electron density $n_e\lesssim 3n_0$, 
then increases {\it slowly} at the moderate electron density $3n_0\lesssim n_e\lesssim 20
n_0$ and finally, decreases slowly when $n_e\gtrsim 20
n_0$. Accordingly, a valley and an
{\it extremely} weak peak appears at low and moderate electron densities,
respectively. Nevertheless, when the Coulomb potential is taken to be
independent on electron density 
in Eq.~(\ref{triplet2}),
it is shown in Fig.~\ref{figyw7} by the green chain curve with circles
 (labeled by $\Delta_{t,C}^m$) that
with the increase of the electron density, $\Delta_{t,C}^m$ first increases when
$n_e\lesssim 3n_0$, then decreases rapidly when $3n_0\lesssim n_e\lesssim 20
n_0$, and finally decreases slowly when $n_e\gtrsim 20
n_0$. One notices that all these
features are very similar to the electron density dependence of the renormalized
singlet order parameter, addressed in detail in
Sec.~\ref{singlet_density}. The only difference is that when $n_e\gtrsim 20
n_0$, the system actually lies in the single-band regime with the electrons
in $E_{{\bf k},+}$-band being efficiently screened (refer to
Sec.~\ref{triplet_momentum}).
 Above features can be understood as follows.

When the electron density is relatively low ($n_e\lesssim 3n_0$), with the increase of the
  electron density, the chemical potential increases slowly, but 
the strength of the Coulomb potential varies rapidly, with a valley
 appearing at the crossover between the non-degenerate and degenerate
  regimes. When the electron density
  is relatively high ($n_e\gtrsim 20n_0$), the effective
  polarization function
  becomes insensitive to the variation of the electron density, and hence with
  the increase of the electron density, the increase of the
  chemical potential influences the triplet pairing and causes the decrease of
  $\Delta_{t}^m$.
 Finally, in the moderate regime
 ($3n_0\lesssim n_e\lesssim 20
n_0$), there exits competition between the Coulomb potential and triplet
pairing, leading to a shallow peak.

Finally, we compare the magnitude of the renormalized singlet and induced triplet order
parameters. In Fig.~\ref{figyw7}, the minimum and maximum values of the renormalized
singlet ($\Delta_{S}^{mi}=\Delta-\Delta_{s}^m$) and triplet ($\Delta_{t}^m$)
 order parameters are plotted by the red solid curve with
circles and blue dashed curve with squares, respectively. Specifically, one observes that when the electron
density $n_e\approx 8n_0$, $\Delta_{S}^{mi}$ and $\Delta_{t}^m$ become
comparable. This provides an ideal condition to observe and distinguish these two order parameters
in experiment. Moreover, with the magnitude comparable to the
  singlet one, the triplet order parameter can provide significant protection to the ground state and is promising to
 cause rich physics
 especially for the elementary excitation.

\section{Summary and discussion}
\label{summary}
In summary, we here demonstrated that the triplet $p$-wave superconductivity can be realized in 
  the strong spin-orbit-coupled QWs in
proximity to $s$-wave superconductor. It is analytically shown that the triplet order parameter is
induced due to the e-e Coulomb and e-p
interactions. Specifically, with the singlet order parameter from
the superconducting proximity effect, not only
can the singlet pairings exist from the proximity-induced order parameter, but
also the
triplet pairings are induced due to the SOC.\cite{Gorkov_Rashba} Then with the
effective e-e interactions, the
singlet order parameter is renormalized from the singlet pairings
 and the triplet order parameter is induced from the
  triplet pairings.  All these can be systematically obtained from the derived effective BdG
 equation, in which the self-energies due to the e-e Coulomb and e-p interactions are proved
 to play the role of the singlet and triplet 
order parameters. Moreover, for the renormalized singlet order
parameter,
 we reveal that it is suppressed because the singlet order parameter
 induced from the repulsive effective e-e interaction is always in opposite sign against the proximity-induced
  one. For the induced triplet order parameter, it is proved
that it is odd in the momentum and is the $p$-wave one ($p_x\pm ip_y$).

We then perform the numerical calculations
  for the renormalized singlet and induced triplet order parameters 
in a specific material, i.e., strong spin-orbit-coupled InSb (100)
 QWs.\cite{material_book,parameter_SOC} In InSb QWs, the calculations show that at
   low temperature, the self-energy contributed
 by the e-p interaction is two orders of magnitude smaller than the one due to
 the e-e Coulomb interaction, and
hence is negligible.  Specifically, for the Coulomb-interaction--induced singlet order parameter,
 it only depends on the
  magnitude of the momentum, which decreases with the
  increase of the energy due to the suppression of the Coulomb
  interaction.
 For the induced triplet order parameter, it depends not only on the magnitude, but also on the angle of the 
  momentum. Specifically, in the module dependence of the momentum, a peak
  shows up at the position determined by
    \begin{equation}
k_c=m^*\alpha+\sqrt{2m^*\mu+m^2\alpha^2}
\end{equation} where the electron energy just corresponds to
    the chemical potential. In the
  angular dependence of momentum, it is revealed that
 the ${\bf d}$-vector of the induced triplet order parameter
is parallel to the effective magnetic field due to the SOC, and hence is protected
by the SOC.\cite{Sigrist_PRL} Finally, it is found that with proper electron
density ($n_e\approx 8\times 10^{14}$~cm$^{-2}$),
  the maximum of the induced triplet order parameter and the minimum of the
  renormalized singlet order parameter are
comparable. This provides an ideal condition to observe and
distinguish these order parameters experimentally.

Finally, we discuss the possibilities to realize the triplet pairing and triplet
  order parameter in other systems including the symmetric (110) and (111) QWs.
 For the (110) symmetric OWs, the Dresselhaus SOC only has the out-of-plane
component, which is expressed as
$H^{(110)}_{\rm soc}=h_z({\bf k})\sigma_z$ with $h_z({\bf k})$ being the effective
magnetic field.\cite{110_18,110_19,110_20,110_21,110_22,Yang,Yu} In this situation, the triplet pairing and
triplet order parameter are exactly zero. For the (111) symmetric QWs,
the Dresselhaus SOC is expressed as $H_{\rm soc}^{(111)}={\bf h}({\bf k})\cdot{\bgreek \sigma}$ with ${\bf h}({\bf
  k})=\big(h_x({\bf k}),h_y({\bf k}),h_z({\bf k})\big)$.\cite{111_18,BoYe}
 The ${\bf d}$-vector of the triplet order parameter is parallel to the in-plane
 components of the 
 SOC, whose strength is influenced by the out-of-plane component of the SOC
 because of its influence on the energy spectra.

\begin{acknowledgments}
This work was supported
 by the National Natural Science Foundation of China under Grant
No. 11334014 and  61411136001, the National Basic Research Program
 of China under Grant No.
2012CB922002 and the Strategic Priority Research Program 
of the Chinese Academy of Sciences under Grant
No. XDB01000000.

\end{acknowledgments}

\begin{appendix}
\section{SINGLE-PARTICLE TUNNELING INDUCED SELF-ENERGY}
\label{AA}
The single-particle tunneling Hamiltonian between the QWs and $s$-wave
  superconductor is given as
\begin{equation}
\hat{H}_{T}=\int d{\bf r}\hat{\Psi}^{\dagger}({\bf
  r})\hat{T}\hat{\tau}_3\hat{\Phi}({\bf r}),
\label{tunneling}
\end{equation}
where $\hat{T}=t$ with $t$ being the element of the tunneling matrix, taken to be real
in this work. One notes that Eq.~(\ref{tunneling}) is widely used in the study
of the quantum nanowire in proximity to superconductor in the study of Majorana zero
 mode.\cite{Dar_Sarma1,Dar_Sarma2,Dar_Sarma3}

Following the derivation in 
Refs.~\onlinecite{Dar_Sarma1,Dar_Sarma2,Hubbard1},
 the self-energy in the Matsubara representation
 due to the single-particle tunneling effect is calculated based
 on Hamiltonian Eq.~(\ref{tunneling}) and is given by 
\begin{equation}
\Sigma_s(\tau_1-\tau_2,{\bf r}_1-{\bf r}_2)=\hat{T}G_S(\tau_1-\tau_2,{\bf r}_1-{\bf r}_2)\hat{T}^{\dagger}.
\label{self_tunneling}
\end{equation}
In Eq.~(\ref{self_tunneling}), ${\bf r}_1$ and ${\bf r}_2$ are 2D in QWs,
which corresponds to the interface between
QWs and superconductors; $G_S(\tau_1-\tau_2,{\bf r}_1-{\bf r}_2)$ is the Green function in the $s$-wave
superconductor, which is defined as
\begin{equation}
G_S(\tau_1-\tau_2,{\bf r}_1-{\bf r}_2)=-\hat{\tau}_3\big\langle
  T_{\tau}\hat{\Phi}(\tau_1,{\bf r}_1)\hat{\Phi}^{\dagger}(\tau_2,{\bf r}_2)\big\rangle.
\end{equation} 
In the frequency-momentum space, the self-energy due to the single-particle tunneling effect
 is further written as\cite{Dar_Sarma1,Dar_Sarma2,Dar_Sarma3,Hubbard1} 
\begin{equation}
\Sigma_s(i\omega_m,{\bf k})=\hat{T}G_S(i\omega_m,{\bf k})\hat{T}^{\dagger},
\label{tunneling_f_p}
\end{equation}
in which $G_S(i\omega_m,{\bf k})=\int \frac{\displaystyle d
  p_z}{\displaystyle 2\pi}G_S(i\omega_m,{\bf p})$ with ${\bf p}=({\bf
  k},p_z)=(k_x,k_y,p_z)$.
  Specifically, in $s$-wave superconductor,
\begin{eqnarray}
\nonumber
&&G_S(i\omega_m, {\bf p})=\frac{1}{(i\omega_m)^2-\zeta_{\bf
    p}^2-|\Delta_0|^2}
\label{Green_super}\\
\nonumber
&&\mbox{}\times\left(
\begin{array}{cccc}
i\omega_m+\zeta_{\bf p} & 0 &0&\Delta_0\\
0&i\omega_m+\zeta_{\bf p}&-\Delta_0&0\\
0&\Delta_0^*&-i\omega_m+\zeta_{\bf p}&0\\
-\Delta_0^*&0&0&-i\omega_m+\zeta_{\bf p}
\end{array}
\right),\\
\end{eqnarray} 
where $\zeta_{\bf p}=\frac{\displaystyle {\bf p}^2}{\displaystyle 2\tilde{m}}-\tilde{\mu}$.
Accordingly, from Eqs.~(\ref{tunneling_f_p}) and (\ref{Green_super}), one obtains
the self-energy due to single-particle tunneling effect in frequency-momentum space
\begin{eqnarray}
\nonumber
&&\Sigma_s(i\omega_m,{\bf k})=t^2\int
\frac{dp_z}{2\pi}\frac{1}{(i\omega_m)^2-\zeta_{\bf p}^2-|\Delta_0|^2}
\label{Self_tunneling_final}\\
\nonumber
&&\mbox{}\times\left(
\begin{array}{cccc}
i\omega_m+\zeta_{\bf p} & 0 &0&\Delta_0\\
0&i\omega_m+\zeta_{\bf p}&-\Delta_0&0\\
0&\Delta_0^*&-i\omega_m+\zeta_{\bf p}&0\\
-\Delta_0^*&0&0&-i\omega_m+\zeta_{\bf p}
\end{array}
\right).\\
\end{eqnarray}

From Eq.~(\ref{Self_tunneling_final}), one observes that
  $\Sigma_s(i\omega_m,{\bf k})$ generally depends on the Matsubara frequency and
  momentum, and hence the real frequency after the analytical continuation
  $i\omega_{m}\rightarrow \omega+i0^{+}$. Specifically, from the effective BdG equation [Eq.~(\ref{effective})], in
 $\Sigma_s(\omega,{\bf k})\hat{\tau}_3$, the diagonal terms only modifies the
 effective mass of the electron and shifts 
zero-energy point of the system, which are neglected in our analysis; whereas
 the off-diagonal terms act as the effective {\it even-frequency} and {\it
   even-momentum} singlet order parameter.\cite{Dar_Sarma1,Dar_Sarma2,Dar_Sarma3,Hubbard1}
 Accordingly, we obtain the tunneling-induced order parameter,
\begin{equation}
\Delta(i\omega_m,{\bf k})=-t^2\int
\frac{dp_z}{2\pi}\frac{\Delta_0}{(i\omega_m)^2-\zeta_{\bf p}^2-|\Delta_0|^2}.
\label{effective_potential}
\end{equation}

\section{COULOMB SCREENING}
\label{BB}
\begin{figure}[ht]
  {\includegraphics[width=8cm]{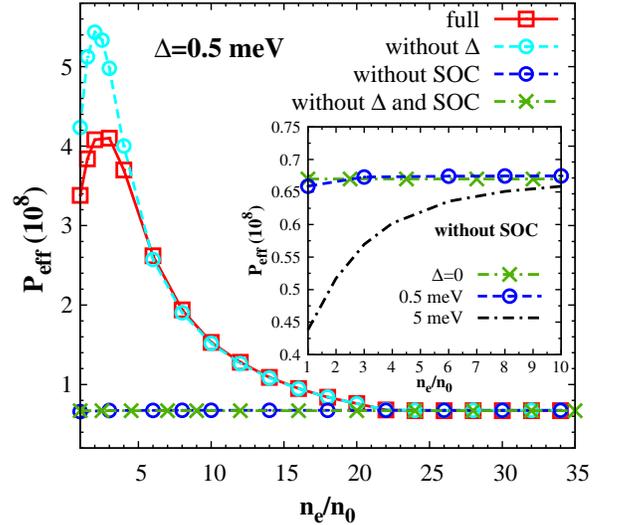}}
  \caption{(Color online) Density dependence of the effective
    polarization function
    $P_{\rm eff}$. The red solid curve with squares represents the full
    calculation with the SOC and $\Delta$ explicitly included. Furthermore,
    the situations without $\Delta$, without the SOC, and without $\Delta$
    and the SOC
    are denoted by the cyan dashed curve with circles, the blue dashed curve with
    circles and the green chain curve with crosses, respectively. The inset
    zooms the electron density dependence of the effective
    polarization function
    when the SOC is set to be zero, with $\Delta=0$ (the green chain curve
    with crosses), 0.5~meV (the blue dashed curve with circles) and 5~meV (the
    black chain curve), respectively.}
  \label{figyw8}
\end{figure}

In this appendix, we present the calculation of the Coulomb screening 
  from the linear response theory,\cite{Fetter,Mahan,Abrikosov} in which both
  the {\it strong} SOC and proximity-induced singlet order parameter are considered explicitly.
 In the Matsubara representation, the dielectric constant in the RPA approximation is calculated by
\begin{equation}
\varepsilon_{\rm RPA}({\bf k},i\omega_n)=1-\tilde{V}_{\bf k}P^{(1)}({\bf k},i\omega_n),
\label{dielectric}
\end{equation}
with $\tilde{V}_{\bf k}$ being the unscreened Coulomb potential. 
In Eq.~(\ref{dielectric}),
\begin{equation}
P^{(1)}({\bf k},i\omega_n)=-\int_0^{\beta}d\tau e^{i\omega_n\tau}\langle
  T_{\tau}\hat{\rho}({\bf k},\tau)\hat{\rho}(-{\bf k},0)\rangle,
\label{dielectric2}
\end{equation}  
in which $\hat{\rho}({\bf k})$ is the density operator.
 Eq.~(\ref{dielectric2}) is further expressed by the $4\times 4$ Green function [Eq.~(\ref{Green_tot})] as
\begin{equation}
P^{(1)}({\bf k},i\omega_n)=\frac{1}{2\beta}\sum_{{\bf
    k}',n'}\mbox{Tr}\Big[G({\bf k}+{\bf k}',\omega_n+\omega_{n'})G({\bf
  k}',\omega_{n'})\Big].
\label{dielectric_four}
\end{equation} 
In our calculation, we focus on the long-wave and static situations, i.e.,
 ${\bf k}\rightarrow 0$ and
$\omega \rightarrow 0$ in Eq.~(\ref{dielectric_four}).
 To reveal the effects of the SOC and proximity-induced singlet order parameter,
 we also calculate the normal case by setting $\alpha$ and/or $\Delta$ to be zero
 in Eq.~(\ref{dielectric_four}). These results are summrized in
 Fig.~\ref{figyw8} in the electron density dependencies of the
 effective polarization function,
 which is defined in Eq.~(\ref{effective_screening}).

In Fig.~\ref{figyw8}, when the SOC and $\Delta$ are explicitly included in
  the calculation, it is shown by the red solid curve with squares that there
exists a peak in the electron density dependence of $P_{\rm eff}$, which appears at the crossover
between the non-degenerate and degenerate regimes. This is in contrast to the
case without the SOC and $\Delta$, referred to as the normal
case. This is shown by the green
chain curve with crosses that when $\alpha$ and
$\Delta$ are set to be zero, $P_{\rm eff}$ becomes insensitive to
the electron density. This insensitivity arises from the fact that when $\alpha$ and
$\Delta$ are set to be zero,
with the electron densities we study here, the system always lies in the degenerate
regime; whereas with the SOC included, the system actually lies in the
non-degenerate regime when
 $n_e\lesssim 3n_0$, as shown in the inset of Fig.~\ref{figyw3}.

To clearly reveal the effects of the SOC and $\Delta$ in the Coulomb screening,
 we further calculate the cases with
only the SOC or $\Delta$ included in Eq.~(\ref{dielectric_four}). In Fig.~\ref{figyw8}, it is shown by the
  cyan dashed curve with circles that when $\Delta$ is set to be zero (hence
  only the SOC is included),
 the effective polarization function also shows a
  peak in the electron density dependence. Specifically, the peak is
  significantly enhanced at the
crossover between the non-degenerate and degenerate regimes 
compared to the full calculation represented by the red solid curve
  with squares. This indicates that
$\Delta$ can suppress the screening effect. Nevertheless, when the SOC is set
to be zero with $\Delta=0.5$~meV, it is shown by the blue dashed curve with
circles that $P_{\rm eff}$ becomes very close to the normal case denoted by the
green chain curve with crosses. Therefore, it is the joint effects of the SOC and
singlet order parameter that cause the efficient suppression of the Coulomb
screening here.
 Actually, the singlet order parameter alone can also
suppress the Coulomb screening. Nevertheless, it is not obvious in the weak
coupling limit when $\Delta$ is much smaller than the Fermi energy,
 but significant when $\Delta$ is large. In the inset of
Fig.~\ref{figyw8}, we show that when $\alpha=0$, compared to the
case with $\Delta=0$, the screening with $\Delta=5$~meV represented by the black chain curve  
shows that $P_{\rm eff}$ is significantly suppressed at low electron density.

\end{appendix}


\begin{thebibliography}{0}
\bibitem{note} By considering the e-e Coulomb interaction for the
electrons between the QWs and $s$-wave superconductor, the singlet
pairings in $s$-wave superconductor can also induce the singlet order
parameter in QWs by providing the self-energy [Eq.~(\ref{effective})].
\end{thebibliography}
\end{document}